\documentclass[showpacs,preprintnumbers,amssymb]{revtex4-2}
\usepackage{graphicx}
\usepackage{dcolumn}
\usepackage{color}
\usepackage{bm}
\usepackage{amsmath}
\usepackage{amssymb}
\usepackage{epsfig}
\usepackage{epsfig}
\usepackage{epstopdf}
\usepackage{multirow}

\def \a{\alpha}
\def \b{\beta}
\def \l{\lambda}
\def \L{\Lambda}

\def \d{\delta}
\def \k{\kappa}
\def \s{\sqrt}
\def \be{\begin{equation}}
\def \ee{\end{equation}}
\def \ben{\begin{eqnarray}}
\def \een{\end{eqnarray}}
\def \o{\omega}

\def \p{\partial}
\def \t{\theta}

\def \G{\bar{G}}

\def \k{\kappa}
\def \R{\bar{R}}
\def \T{\bar{T}}
\def \sq{\bar{\square}}
\def \ga{\Gamma}
\def \sg{\sigma}

\begin{document}
	
\title{ $f(\bar{R}, L(X))$-gravity in the context of dark energy with power law expansion and energy conditions}

\author{Goutam Manna}
\altaffiliation{goutammanna.pkc@gmail.com}
\author{Arijit Panda$^a$}
\altaffiliation{arijitpanda260195@gmail.com}
\author{Aninda Karmakar}
\altaffiliation{aninda134@gmail.com}
\affiliation{Department of Physics, Prabhat Kumar College, Contai, Purba Medinipur-721404, India}
\author{Saibal Ray }
\altaffiliation{saibal@associates.iucaa.in}
\affiliation{Department of Physics, Government College of Engineering \& Ceramic Technology, 73, Abinash Chandra Banerjee Lane, Kolkata-700 010, India}
\author{Md. Rabiul Islam}
\altaffiliation{rabi76.mri@gmail.com}
\affiliation{$^a$ Department of Physics, Raiganj University, Raiganj, Uttar Dinajpur-733 134, West Bengal, India.}

\begin{abstract}
The motto of this work is to generate a general formalism of $f(\bar{R}, L(X))-$gravity in the context of dark energy under the framework of the {\bf K-}essence emergent geometry with the Dirac-Born-Infeld (DBI) variety of action, where $\bar{R}$ is the familiar Ricci scalar,  $L(X)$ is the DBI type non-canonical Lagrangian with $X={1\over 2}g^{\mu\nu}\nabla_{\mu}\phi\nabla_{\nu}\phi$ and $\phi$ is the {\bf K-}essence scalar field. The emergent gravity metric $\G_{\mu\nu}$ and the well known gravitational metric $g_{\mu\nu}$ are not conformally equivalent. We have constructed a modified field equation using the metric formalism in $f(\bar{R}, L(X))$-gravity incorporating the corresponding Friedmann equations in the framework of the background gravitational metric which is of Friedmann-Lema{\^i}tre-Robertson-Walker (FLRW) type. The solution of modified Friedmann equations have been deduced for the  specific choice of $f(\bar{R}, L(X))$, which is of Starobinsky-type, using power law expansion method. The consistency of the model with the accelerating phase of the Universe has been shown, when we restrict ourselves to consider the value of the dark energy density, as $\dot\phi^{2}=\frac{8}{9}=0.888 <1$, which indicates that the present Universe is dark energy dominated. Graphical plots for the energy density ($\rho$), pressure ($p$) and equation of state parameter ($\o$) w.r.t. time ($t$) based on parametric values are interestingly consistent with the dark energy domination and hence  accelerating features. We also put some light on the corresponding energy conditions and constraints of the $f(\bar{R}, L(X))$ theory with one basic example.


\end{abstract}

\maketitle

\section{Introduction}
Weyl in 1919~\cite{weyl} and Eddington in 1923~\cite{eddin} included higher order constant terms concerned with the curvature in the action of General Relativity firstly given by Albert Einstein. We can quantize GR by the non-renormalizability of the theory in a conventional way. Utiyama and  De Witt~\cite{uti} in 1962 showed that we can renormalize GR in one loop if, the Einstein-Hilbert (EH) action is constructed with higher-order curvature terms. In addition, when quantum corrections or string theory entered in the scenario, the action of effective gravitational field with low energy enrols higher-order curvature terms~\cite{birrel}. Being inspired by this argument, scientists have been tried to modify the EH action in applied higher order theories of gravity. 

Therefore, higher-orders of Ricci scalar, should be incorporate in the action. We should keep in mind that, the modified theories of gravity are important at scales only, which are in the range of the Planck scale, i.e., in the newborn universe or at the black hole (BH) singularities, e.g., inflation due to curvature case~\cite{staro}. Sotiriou and Felice \cite{sotiriou,felice} made a review of the well-known ``$f(R)$ theory of gravity''. This $f(R)$ gravity was reconstructed by Dunsby et al.~\cite{dunsby} taking the background as Friedman-Lemaitre-Robertson-Walker (FLRW) type. Particularly, the authors predicted the only Lagrangian with a real value for which $f(R)$ can generate a true $\L CDM$ expansion for dust-like matter-filled universe and is equivalent to the EH Lagrangian with a cosmological constant which is positive. A general formalism was proposed by Mukherjee and Banerjee~\cite{mukherjee} by which we can observe the later time dynamics of the universe for a given analytic model of $f(R)$ gravity, considering the cold dark matter. The outcomes of energy conditions of the $f(R)$ gravity for various cosmological cases have also been investigated in~\cite{atazadeh,santos,capoz,wang1,berg}. The energy conditions using the Raychaudhuri equations in expanding universe were also studied in ~\cite{alba1,alba2,krori}.

Harko et al.~\cite{harko} considered the action as an arbitrary function of not only the Ricci scalar ($R$) but also the matter Lagrangian ($L_{m}$).  With the help of the metric formalism the authors achieved the gravitational field equations of $f(R,L_{m})$ gravity, along with the equations of motion (EOM) of test particles. Later on, Wang et al.~\cite{wang2} have achieved the general energy conditions of the $f(R,L_{m})$ gravity. In~\cite{Goheer1,Goheer2,Singh}, the authors have described the power-law cosmic expansion in higher derivative gravity. Later, Harko et. al., included the trace of the Stress-energy tensor in the action along with the Ricci scalar and developed a new theory, named as `$f(R,T)$ theory' \cite{Harko3}. This type of modified gravity gain massive popularity in recent times to study various cosmological phenomena. 

The development of the {\bf K-}essence model was carried out in~\cite{Picon1,Picon2,babi1,babi2,babi3,babi4,babi5,scherrer1,scherrer2}. On the basis of the Dirac-Born-Infeld (DBI) model~\cite{born1,born2,born3}, Manna et. al.~\cite{gm1,gm2,gm3} has derived a {\bf K-}essence emergent gravity metric $\bar G_{\mu\nu}$, which has different significance than the usual gravitational metric, $g_{\mu\nu}$. The {\bf K-}essence model~\cite{Picon1,Picon2,babi1,babi2,babi3,babi4,babi5,scherrer1,scherrer2} is basically a scalar field theory, where the kinetic energy of the {\bf K-}essence field rules over the potential energy of the field. The Lagrangian corresponding to the {\bf K-}essence field is of non-canonical type. The difference between {\bf K-}essence theory which incorporates non-canonical kinetic parts and canonical relativistic field theories lies in the non-trivial dynamical solutions of the {\bf K-}essence EOM. Besides spontaneous breaking of Lorentz invariance, it changes the perturbed metric near the solutions also. Hence, these perturbations moves along the formal {\it emergent or analogue} curved space-time with the perturbed metric. The Lagrangian for {\bf K-}essence model can be written as, $L=-V(\phi)F(X)$, where $\phi$ is the {\bf K-}essence scalar field and $X=\frac{1}{2}g^{\mu\nu}\nabla_{\mu}\phi\nabla_{\nu}\phi$.

A modified version of $f(R)$ gravity was also studied by Nojiri et al.~\cite{nojiri}, in which the authors have considered the higher order kinetic terms of a scalar field, which was incorporated in vacuum $f(R)$ gravity's action part. Actually, the authors included a general class of the {\bf K-}essence Lagrangian $G(X)$ with the action of vacuum  $f(R)$ gravity.  In the background of slow-roll approximation the authors have investigated the inflationary sides of their theory. Odintsov et. al.~\cite{odin} had analysed the consequences of {\bf K}-essence geometry in the $f(R)$ gravity when cold dark matter and radiation is present. Perfect fluids with the similar model was considered in~\cite{nojiri}. Several cosmological quantities, such as the dark energy equation of state parameter ($\o$), the dark energy density parameter ($\Omega_d$), and some state finder quantities were also some of their findings. Recently, Oikonomou et al.~\cite{oiko} have discussed the phase space of a simple {\bf K-}essence $f (R)$ gravity theory. 

At this juncture, let us discuss the importance of the {\bf K-}essence theory with specific choice of DBI type Lagrangian:  It is now unavoidable to admit the acceleration of the universe after analysing the observations of Large-Scale Structure, type Ia Supernovae's observations and measurements of anisotropy of Cosmic Microwave Background ~\cite{Bahcall}. It has also been acknowledged that our universe is now dominated by a component named dark energy which poses negative pressure. Scientists proposed cosmological constant or vacuum density to be the candidate for such an exotic component of the universe. But the barrier of cosmological coincidence problem created hurdles for us and raised the question that {``\it why does the strange dark energy component possesses a tiny energy density $(O(meV^4  ))$ compared to the simple expectation based on quantum field theory"}. Additionally, {``\it the occurrence of the acceleration at such a late stage of evolution"} continuously poked cosmologists for better theories. The problem with most of the dark energy models (take cosmological constant as example) are that they require {\it extraordinary fine tuning} of the initial energy density which is of the orders of $100$ or more smaller than the initial matter-energy density.

Here comes the new class model with scalar field having amazing dynamic properties that may avoid the long awaited fine-tuning problem, known as {\bf K-}essence theory~\cite{Picon1}. The most promising feature of this model is that it brings the negative pressure from its {\it nonlinear kinetic energy} of the scalar field. It has already been investigated that for a wide class of theories, there exist attractor solution~\cite{Picon2,Kang} in which the scalar field propagates with different evolution speed to achieve the required equation of state of the {\bf K-}essence theory in different epoch with changing equation of state of the background. While our universe was going through the epoch of radiation domination, {\bf K-}essence field was driven to be superior and by imitating the equation of state (EOS) of the radiation it keeps the ratio of {\bf K-}essence field and radiation density constant. At the time of dust domination the {\bf K-}essence theory failed to mimic the EOS of dust-like phase for its dynamical characteristics. It also decreased its energy value quickly by a large amount of magnitude and settled upon at a constant value. Afterwards the field outgrew the matter density and took the universe into cosmic acceleration at a time, roughly corresponding to the current age of the universe. At last, the EOS of {\bf K-}essence theory slowly reclines to an value between the range $0$ and $-1$.
 
The well-known {\it Quintessence Trackers models}~\cite{Caldwell,Frieman,Peebles,Ratra,Zlatev1,Zlatev2,Zlatev3} almost give the same result as given by the {\bf K-}essence theory but with one problem. It is good that it can mimic the equation of state of matter and radiation of background EOS, but it requires an adjustable parameter which need to be finely tuned to get the preferable energy density which can produce the negative pressure at present age. 

{\bf K-}essence theory is different in the sense that it traces the background energy density when the universe was in radiation epoch only. The sharp transition of positive pressure to negative pressure at the matter-radiation equality occurs automatically with the help of its dynamics. {\bf K-}essence theory was unable to dominate before matter-radiation equality as it was busy to track the radiation background. Since the energy density inevitably drops to a very small value at the transition to dust domination, it was also impossible to dominate immediately just after the dust domination. On the other hand, as the matter density drops more rapidly than the energy density with expanding universe, {\bf K-}essence field came into control at the age roughly around the current epoch. Thus, the whole problem of Cosmic Coincidence Problem (i.e., why we live in the era of dark matter and dark energy density's equality) vanishes to the fact that we came to observe the universe at the right time after matter-radiation equality. 

{\bf K-}essence models also ensures the production of dark energy component where the sound speed ($c_{s}$) does not exceed the light speed. There exist a difference between these models and scalar field quintessence models from the observational background with a canonical kinetic term (for which $c_{s} = 1$), and this may one of the ways to reduce cosmic microwave background (CMB) fluctuations measured on large angular scales \cite{eric,dedeo,bean}. Though there may be some stages where the fluctuations of the field can propagate superluminally ($c_{s}>1$)~\cite{babi3,babi4,Bonvin}.

Some cosmological behaviours and the stability of the {\bf K-}essence model in FLRW spacetime has been studied by Yang et al.~\cite{Yang}.
Some opposite results have been obtained for small sounds speed of scalar perturbations which implies clustering of dark energy and increase of cosmological perturbations \cite{Sawicki,Kunz}.

Historically, Born and Infeld~\cite{born1} introduced a non-canonical kinetic theory to overcome the infinite self-energy of the electron. Some more non-canonical theories was also studied in literatures like~\cite{born2,born3}. The studies~\cite{Linde,Albrecht,Dvali,Kachru,Alishahiha,Silverstein1,Chen1,Weinberg,Chen2} of string theory, brane cosmology, D-branes, etc. have also used the DBI type non-canonical Lagrangian.

Motivated by the above said importance of the {\bf K-}essence theory which prescribes a way to investigate the effects of the presence of dark energy component in cosmological framework, in this paper we study the $f(R)-$gravity in the context of {\bf K-}essence emergent gravity, i.e., dark energy in a general manner. We have made the generalization of the $f(\R,L(X))$ theory with the help of the metric formalism, where $\R$ is the Ricci scalar of the {\bf K}-essence geometry and $L(X)$ is the DBI type non-canonical Lagrangian. Panda et. al. \cite{Panda} have modified the $f(R,T)$ theory in the context of dark energy using the {\bf K-}essence model. The process of studying of these two papers are different from the begining i.e., consideration of actions are different. 

Also, we have calculated the energy conditions and modified Friedman equations in $f(\R,L(X))$ gravity, where we have considered the flat FLRW type metric as the background gravitational metric. The modified field equation, Friedmann equations and energy conditions for the new $f(\R,L(X))$ gravity theory are different from the usual $f(R,L_{m})$~\cite{harko} and $f(R)$~\cite{sotiriou} gravity theories. Also, we have solved the modified Friedmann equations using power law cosmic expansion method.

The paper is organized as follows:  In Sec. 2, we have briefly discussed about the {\bf K-}essence emergent geometry with the help of the following works~\cite{gm1,gm2,gm3,babi1,babi2,babi3,babi4,babi5}. In Sec. 3, we have formulated the $f(\R,L(X))$ gravity in the context of the {\bf K}-essence emergent geometry. We also have derived the modified field equations and the condition of the requirement of the conservation of the energy-momentum tensor in the $f(\R,L(X))$ gravity.  The modified Friedmann equations were brought into limelight in Sec. 4, considering the background gravitational metric as flat FLRW and the {\bf K-}essence scalar field as a function of time only. The solution of Friedmann equations have been solved for specific choice of $f(\R,L(X))$ using power law method in Sec. 5, whereas in Sec. 6 we develop the energy conditions and constraints of the $f(\R,L(X))$ gravity with example. The last section Sec. 7 contains some general discussion and crucial conclusion of our work. Also, we have briefly discussed the $f(R)$-gravity and $f(R,L_{m})-$gravity and corresponding energy conditions~\cite{atazadeh,santos,capoz,wang1,wang2,berg,carroll} in  {\bf Appendix}.

\section{{\bf K}-essence theory: a glimpse of background and development}
In this section, we will discuss about the development of the modified metric corresponding to the emergent spacetime which is related with the general background geometry along with a very general {\bf K}-essence scalar field. The {\bf K}-essence scalar field $\phi$, has action~\cite{babi1,babi2,babi3,babi4,babi5}
\ben
S_{k}[\phi,g_{\mu\nu}]= \int d^{4}x {\sqrt -g} L(X,\phi),
\label{14}
\een
which has a minimal coupling with the background space-time metric $g_{\mu\nu}$ and $X={1\over 2}g^{\mu\nu}\nabla_{\mu}\phi\nabla_{\nu}\phi$ represents the canonical kinetic term.
The energy-momentum tensor is
\ben
T_{\mu\nu}&\equiv& {-2\over \sqrt {-g}}{\delta S_{k}\over \delta g^{\mu\nu}}=-2\frac{\p L}{\p g^{\mu\nu}}+g_{\mu\nu}L
\nonumber\\&&=-L_{X}\nabla_{\mu}\phi\nabla_{\nu}\phi+g_{\mu\nu}L,
\label{15}
\een
with $L_{\mathrm X}= {dL\over dX},~L_{\mathrm XX}= {d^{2}L\over dX^{2}},~L_{\mathrm\phi}={dL\over d\phi}$ and the symbol $\nabla_{\mu}$ stands for the covariant derivative with respect to the gravitational metric $g_{\mu\nu}$.

The EOM of scalar field is
\ben
-{1\over \sqrt {-g}}{\delta S_{k}\over \delta \phi}= G^{\mu\nu}\nabla_{\mu}\nabla_{\nu}\phi +2XL_{X\phi}-L_{\phi}=0,
\label{16}
\een
where  
\ben
G^{\mu\nu}\equiv \frac{c_{s}}{L_{X}^{2}}[L_{X} g^{\mu\nu} + L_{XX} \nabla ^{\mu}\phi\nabla^{\nu}\phi],
\label{17}
\een
with $1+ {2X  L_{XX}\over L_{X}} > 0$ and $c_s^{2}(X,\phi)\equiv{(1+2X{L_{XX}\over L_{X}})^{-1}}$.

The inverse metric, $G^{\mu\nu}$ can be written in the form of
\ben G_{\mu\nu}={L_{X}\over c_{s}}[g_{\mu\nu}-{c_{s}^{2}}{L_{XX}\over L_{X}}\nabla_{\mu}\phi\nabla_{\nu}\phi].
\label{18}
\een

Applying a conformal transformation further~\cite{gm1,gm2} $\bar G_{\mu\nu}\equiv {c_{s}\over L_{X}}G_{\mu\nu}$ gives
\ben 
\bar G_{\mu\nu}
={g_{\mu\nu}-{{L_{XX}}\over {L_{X}+2XL_{XX}}}\nabla_{\mu}\phi\nabla_{\nu}\phi}.
\label{19}
\een	

Using Eq. (\ref{15}), the effective emergent metrics (\ref{19}) can be written as~\cite{babi3,babi4}
\ben
\bar G_{\mu\nu}=\left(1-\frac{LL_{XX}}{L_{X}(L_{X}+2XL_{XX})}\right)g_{\mu\nu}+\frac{L_{XX}}{L_{X}(L_{X}+2XL_{XX})}T_{\mu\nu}. \label{20}
\een

We should always keep it in mind that, $L_{X}\neq 0$ as $c_{s}^{2}$ is positive definite and only then Eqs. (\ref{14}) -- (\ref{17}) will have a meaningful physics.

Obviously, If $\phi$ has a non-trivial space-time configuration, then usually the emergent metric $\bar G_{\mu\nu}$ is not conformally equivalent to $g_{\mu\nu}$. So $\phi$ has dissimilar characteristics as of the canonical scalar fields with the locally defined causal structure. Further, {\it if there is no explicit dependency of $L$ on $\phi$,} the reformed EOM (\ref{16}) becomes
\ben
-{1\over \sqrt {-g}}{\delta S_{k}\over \delta \phi}
= \bar G^{\mu\nu}\nabla_{\mu}\nabla_{\nu}\phi=0.
\label{21}
\een

The authors \cite{born1,born2,born3,babi4,babi5,gm1,gm2,gm3,gm4} take the Dirac-Born-Infeld (DBI) type non-canonical Lagrangian as 
$L(X,\phi)=V(\phi)[ 1-\sqrt{1-2X}]$,
where $V(\phi)$ is a constant potential and kinetic energy of the {\bf K-}essence scalar field is very very greater than the potential part of the Lagrangian. In this article, we choose the DBI type non-canonical Lagrangian to be an explicit function of $X$ only as $L(X,\phi)\simeq L(X)$ \cite{Mukohyama} without any loss of generality and dimensionality since in the {\bf K-}essence theory the kinetic energy is dominates over the potential energy of the system. Therefore, we can write the Lagrangian as
\ben
L(X,\phi)\simeq L(X)= V\big[1-\sqrt{1-2X}\big],
\label{22}
\een
where $V$ is a constant potential term.
Then $c_{s}^{2}(X,\phi)=1-2X$ and hence the effective emergent metric (\ref{19}) turns out to be
\ben
\bar G_{\mu\nu}= g_{\mu\nu} - \nabla_{\mu}\phi\nabla_{\nu}\phi\equiv g_{\mu\nu} - \p_{\mu}\phi\p_{\nu}\phi,
\label{23}
\een
since $\phi$ is a scalar. 

Using Eq. (\ref{15}), Eq. (\ref{23}) can be rewritten as $T_{\mu\nu}$ and $\nabla_{\mu}\phi$, as 
\ben
\bar G_{\mu\nu}L =L_{X}\nabla_{\mu}\phi\nabla_{\nu}\phi+T_{\mu\nu}-L\nabla_{\mu}\phi\nabla_{\nu}\phi. 
\label{24}
\een

Following~\cite{wald,gm1} the relation between the new Christoffel symbols and the old ones are  
\ben
\bar\Gamma ^{\alpha}_{\mu\nu} 
&=&\Gamma ^{\alpha}_{\mu\nu} + (1-2X)^{-1/2}\bar{G}^{\alpha\gamma}[\bar{G}_{\mu\gamma}\partial_{\nu}(1-2X)^{1/2}
+\bar{G}_{\nu\gamma}\partial_{\mu}(1-2X)^{1/2}-\bar{G}_{\mu\nu}\partial_{\gamma}(1-2X)^{1/2}]\nonumber\\
&=&\Gamma ^{\alpha}_{\mu\nu} -\frac {1}{2(1-2X)}[\delta^{\alpha}_{\mu}\partial_{\nu}X
+ \delta^{\alpha}_{\mu}\partial_{\nu}X].\label{25}
\een
 
Therefore, using the new Christoffel connections  $\bar\Gamma$ the geodesic equation for the {\bf K}-essence becomes
\ben
\frac {d^{2}x^{\alpha}}{d\l^{2}} +  \bar\Gamma ^{\alpha}_{\mu\nu}\frac {dx^{\mu}}{d\l}\frac {dx^{\nu}}{d\l}=0, \label{26}
\een
where $\l$ is an affine parameter.

Now introducing the covariant derivative $D_{\mu}$~\cite{babi3,babi4}, corresponding to the emergent metric $\bar G_{\mu\nu}$ $(D_{\a}\bar G^{\a\b}=0)$ as 
\ben
D_{\mu}A_{\nu}=\p_{\mu} A_{\nu}-\bar \Gamma^{\l}_{\mu\nu}A_{\l},
\label{27}
\een
and the inverse of the emergent metric is $\bar G^{\mu\nu}$ such that $\bar G_{\mu\l}\bar G^{\l\nu}=\delta^{\nu}_{\mu}$.

Ultimately, the ``emergent'' Einstein's equation becomes
\ben
\bar{E}_{\mu\nu}=\R_{\mu\nu}-\frac{1}{2}\bar{G}_{\mu\nu}\R=\k T_{\mu\nu}, \label{28}
\een
where $\k=8\pi G$ is a constant, $\R_{\mu\nu}$ is emergent gravity's Ricci tensor and $\R~ (=\R_{\mu\nu}\bar{G}^{\mu\nu})$ is the Ricci scalar of the emergent space-time.

\section{$f(\R,L(X))$ Gravity in the context of {\bf K}-essence emergent space-time}
 We are now considering the action of the modified gravity in the context of the {\bf K}-essence emergent space-time, which takes the following form ($\k=1$)
\ben
S= \int d^{4}x \sqrt{-\bar{G}}~ f(\R,L(X)),
\label{29}
\een
where $f(\R,L(X))$ is an arbitrary function of the Ricci scalar $\R$ and the non-canonical Lagrangian density $L(X)$ corresponding to the {\bf K-}essence theory, and $$\sqrt{-\G}=\sqrt{-det({\G_{\mu\nu}})}.$$

Based on~\cite{harko},  varying the action $S$ with respect to the {\bf K}-essence emergent gravity metric $\bar{G}^{\mu\nu}$ we obtain
\ben
\delta S = \int \Big[f_{\R}(\R, L)\delta \R + f_{L}(\R, L)\frac{\delta L}{\delta \bar{G}^{\mu\nu}}\delta \bar{G}^{\mu\nu}- \frac{1}{2}\bar{G}_{\mu\nu}f(\R, L)\delta \bar{G}^{\mu\nu} \Big] \times\sqrt{-\bar{G}}~d^{4}x,
\label{30}
\een
where we have denoted $f_{\R}(\R, L)=\frac{\p{f(\R, L)}}{\p \R}$ and $f_{L}(\R, L)=\frac{\p{f(\R, L)}}{\p L}$. 

Now we obtain the variation of the Ricci scalar for the {\bf K}-essence emergent gravity metric
\ben
\delta \R = \delta (\R_{\mu\nu}\bar{G}^{\mu\nu}) = \delta \R_{\mu\nu}\bar{G}^{\mu\nu} + \R_{\mu\nu}\delta \bar{G}^{\mu\nu}\nonumber\\
=\R_{\mu\nu}\delta \bar{G}^{\mu\nu}+\G^{\mu\nu}(D_{\l}\delta\bar{\Gamma}^{\l}_{\mu\nu}-D_{\nu}\delta\bar{\Gamma}^{\l}_{\mu\l}), 
\label{31}
\een
where
\ben
\R_{\mu\nu} = \partial_{\mu}\bar{\Gamma}^{\alpha}_{\alpha\nu} - \partial_{\alpha}\bar{\Gamma}^{\alpha}_{\mu\nu} + \bar{\Gamma}^{\alpha}_{\beta\mu}\bar{\Gamma}^{\beta}_{\alpha\nu} - \bar{\Gamma}^{\alpha}_{\alpha\beta}\bar{\Gamma}^{\beta}_{\mu\nu},
\label{32}
\een
\ben
\bar{\Gamma}^{\alpha}_{\mu\nu} = \frac{1}{2}\bar{G}^{\alpha\beta}[\partial_{\mu}\bar{G}_{\beta\nu} + \partial_{\nu}\bar{G}_{\mu\beta} - \partial_{\beta}\bar{G}_{\mu\nu}],
\label{33}
\een
and the variation of $\delta\bar{\Gamma}^{\l}_{\mu\nu}$ is
\ben
\delta\bar{\Gamma}^{\l}_{\mu\nu}=\frac{1}{2}\G^{\l\a}[D_{\mu}\d\G_{\nu\a}+D_{\nu}\d\G_{\mu\a}-D_{\a}\d\G_{\mu\nu}].
\label{34}
\een

Thus, the expression for the variation of Ricci scalar $\d\R$ is
\ben
\d\R=\R_{\mu\nu}\d\G^{\mu\nu}+\G_{\mu\nu}D_{\a}D^{\a}\d\G^{\mu\nu}-D_{\mu}D_{\nu}\d\G^{\mu\nu}.
\label{35}
\een
 
 Therefore, variation of the action (\ref{30}) is
 \ben
 \delta S&=& \int \Big[f_{\R}(\R, L)\R_{\mu\nu}\delta \bar{G}^{\mu\nu} + f_{\R}(\R, L)\bar{G}_{\mu\nu}D_{\a}D^{\a}\delta \bar{G}^{\mu\nu}\nonumber\\&&- f_{\R}(\R, L)D_{\mu}D_{\nu}\delta \bar{G}^{\mu\nu} + f_{L}(\R, L)\frac{\delta L}{\delta \bar{G}^{\mu\nu}}\delta \bar{G}^{\mu\nu}\nonumber\\&&- \frac{1}{2}\bar{G}_{\mu\nu}f(\R, L)\delta \bar{G}^{\mu\nu}\Big]\sqrt{-\bar{G}}d^{4}x.~~~~~ \label{36}
 \een
 
After partially integrating second and third terms of the above Eq. (\ref{36}), we get
\ben
\delta S &=& \int \Big[f_{\R}(\R, L)\R_{\mu\nu} + \bar{G}_{\mu\nu}D_{\mu}D^{\mu}f_{\R}(\R, L) - D_{\mu}D_{\nu}f_{\R}(\R, L)  + f_{L}(\R, L)\frac{\delta L}{\delta \bar{G}^{\mu\nu}} \nonumber\\&&- \frac{1}{2}\bar{G}_{\mu\nu}f(\R, L)\Big]\delta \bar{G}^{\mu\nu}\sqrt{-\bar{G}}d^{4}x. \label{37}
\een

Therefore, using principle of least action, i.e. $\delta S = 0$, we have the modified field equation for $f(\R, L(X))$ theory
\ben
&&f_{\R}(\R, L)\R_{\mu\nu} + \bar{G}_{\mu\nu}D_{\a}D^{\a}f_{\R}(\R, L) - D_{\mu}D_{\nu}f_{\R}(\R, L) - \frac{1}{2}\bar{G}_{\mu\nu}f(\R, L) + f_{L}(\R, L)\frac{\delta L}{\delta \bar{G}^{\mu\nu}} = 0.
\label{38}
\een

Now, we evaluate the term $\frac{\delta L}{\delta \bar{G}^{\mu\nu}}$ as
\ben
\frac{\delta L}{\delta \bar{G}^{\mu\nu}}=\frac{\d L}{\d X}\frac{\d X}{\d g^{\mu\nu}}\frac{\d g^{\mu\nu}}{\d\G^{\mu\nu}}=\frac{1}{2}L_{X}D_{\mu}\phi D_{\nu}\phi(1+D_{\a}\phi D^{\a}\phi),~~~~
\label{39}
\een
since for the scalar field $\nabla_{\mu}\phi\equiv\p_{\mu}\phi\equiv D_{\mu}\phi$.

Using Eqs. (\ref{24}), (\ref{38}) and (\ref{39}), we obtain the expression for {\it modified field equation for the $f(\R, L(X))$ theory} in terms of $T_{\mu\nu}$ as
\ben
&&f_{\R}(\R, L)\R_{\mu\nu} +\left(\bar{G}_{\mu\nu}\sq - D_{\mu}D_{\nu}\right)f_{\R}(\R, L)- \frac{1}{2}\left
[f(\R, L)-Lf_{L}(\R, L)(1+D_{\a}\phi D^{\a}\phi)\right]\bar{G}_{\mu\nu}\nonumber\\ &&+\frac{1}{2}Lf_{L}(\R, L)D_{\mu}\phi D_{\nu}\phi\left[1+D_{\a}\phi D^{\a}\phi\right]=\frac{1}{2}f_{L}(\R, L)T_{\mu\nu}\left[1+D_{\a}\phi D^{\a}\phi\right]=\frac{1}{2}f_{L}(\R, L)\T_{\mu\nu},~~~~\label{40}
\een
where $\sq=D_{\mu}D^{\mu}$ and $\T_{\mu\nu}=T_{\mu\nu}[1+D_{\a}\phi D^{\a}\phi]$. 

The above Eq. (\ref{40}) is different from the usual Eq. (\ref{6}) of $f(R,L_{m})$ theory in the presence of the {\bf K}-essence scalar field (vide {\bf Appendix}). If we consider the emergent gravity metric $\G_{\mu\nu}$ is conformally equivalent to the gravitational metric $g_{\mu\nu}$ and $L$ can be matter Lagrangian then we get back to the usual $f(R,L_{m})$ theory in the absence of the {\bf K}-essence scalar field. Also, if we consider $f(\R,L(X))\equiv f(R,L_{m})\equiv \frac{1}{2}R+L_{m}$, i.e., the Hilbert-Einstein Lagrangian form, then from (\ref{40}), we lead to the standard Einstein field equation $R_{\mu\nu}-\frac{1}{2}g_{\mu\nu}R=T_{\mu\nu}$.

Contracting the above field equation (\ref{40}) with $\G^{\mu\nu}$, we have the modified trace equation for the $f(\R, L(X))$ theory is
\ben
&&f_{\R}(\R, L)\R + 3\sq f_{\R}(\R, L)-2\Big[f(\R, L)-f_{L}(\R, L)L(1+D_{\a}\phi D^{\a}\phi)\Big]+\frac{1}{2}Lf_{L}(\R, L)D_{\mu}\phi D^{\mu}\phi[1+D_{\a}\phi D^{\a}\phi]\nonumber\\&&=\frac{1}{2}f_{L}(\R, L)\T, \label{41}~~~~~~
\een
where  $\T=\T^{\mu}_{\mu}$ is trace of the energy-momentum tensor. 

On subtracting Eq. $(\ref{41})\times \G_{\mu\nu}$ from Eq. $(\ref{40})\times 3$, we get
\ben
&&f_{\R}(\R, L)(\R_{\mu\nu}-\frac{1}{3}\G_{\mu\nu}\R)+\frac{1}{6}\G_{\mu\nu}[f(\R, L)-Lf_{L}(\R, L)(1+D_{\a}\phi D^{\a}\phi)]+\frac{1}{3}Lf_{L}(\R, L)D_{\mu}\phi D_{\nu}\phi[1+D_{\a}\phi D^{\a}\phi]\nonumber\\&&=\frac{1}{2}f_{L}(\R, L)(\T_{\mu\nu}-\frac{1}{3}\G_{\mu\nu}\T)+D_{\mu}D_{\nu}f_{\R}(\R, L),
\label{42}
\een
which is an another form of the modified field equation in the presence of {\bf K}-essence scalar field $\phi$.

By taking covariant divergence with respect to $D^{\mu}$ of Eq. (\ref{40}), we have
\ben
&&D^{\mu}\Big[f_{\R}(\R, L)\Big]\R_{\mu\nu}-(\sq D_{\nu}-D_{\nu}\sq)f_{\R}(\R, L)+f_{\R}(\R, L)D^{\mu}\Big(\R_{\mu\nu}-\frac{1}{2}\G_{\mu\nu}\R\Big)+\frac{1}{2}\G_{\mu\nu}f_{\R}(\R,L)D^{\mu}(\R)\nonumber\\&&-\frac{1}{2}D^{\mu}\Big[f(\R,L)\Big]\G_{\mu\nu}+\frac{1}{2}D^{\mu}\Big[Lf_{L}(\R, L)(1+D_{\a}\phi D^{\a}\phi)\Big]\G_{\mu\nu} +\frac{1}{2}D^{\mu}\Big[Lf_{L}(\R, L)D_{\mu}\phi D_{\nu}\phi(1+D_{\a}\phi D^{\a}\phi)\Big]\nonumber\\&&=\frac{1}{2}D^{\mu}\Big[f_{L}(\R, L)\T_{\mu\nu}\Big]. \label{43}
\een

 Using the identities on purely geometrical grounds~\cite{misner,koiv,harko1} $D^{\mu}(\R_{\mu\nu}-\frac{1}{2}\G_{\mu\nu}\R)=D^{\mu}\bar{E}_{\mu\nu}=0$;~\\ $D^{\mu}[f_{\R}(\R, L)]\R_{\mu\nu}=(\sq D_{\nu}-D_{\nu}\sq)f_{\R}(\R, L)$ and also using Eqs. (\ref{24}) and (\ref{39}), the above Eq. (\ref{43}) becomes
\ben
&&D^{\mu}[f_{L}(\R, L)\T_{\mu\nu}]=-f_{L}(\R, L)D^{\mu}[\G_{\mu\nu}L] +D^{\mu}\left[Lf_{L}(\R, L)(1+D_{\a}\phi D^{\a}\phi)(\G_{\mu\nu}+D_{\mu}\phi D_{\nu}\phi)\right]\nonumber\\
&&\Rightarrow D^{\mu}\T_{\mu\nu}=D^{\mu}~ln[f_{L}(\R,L)]\times\left[L_{X}D_{\mu}\phi D_{\nu}\phi (1+D_{\a}\phi D^{\a}\phi)\right]+D^{\mu}\left[L D_{\mu}\phi D_{\nu}\phi (1+D_{\a}\phi D^{\a}\phi)+L\G_{\mu\nu}D_{\a}\phi D^{\a}\phi\right]\nonumber\\
&&\Rightarrow D^{\mu}\T_{\mu\nu}=2D^{\mu}~ln[f_{L}(\R,L)]\frac{\delta L}{\delta \G^{\mu\nu}}+D^{\mu}\left[L D_{\mu}\phi D_{\nu}\phi (1+D_{\a}\phi D^{\a}\phi)+L\G_{\mu\nu}D_{\a}\phi D^{\a}\phi\right].
\label{44}
\een

Thus, the requirement of the conservation of the energy-momentum tensor $(D^{\mu}\T_{\mu\nu}=0)$ for the {\bf K}-essence
Lagrangian, gives an effective functional relation as
\ben
2D^{\mu}~ln[f_{L}(\R,L)]\frac{\delta L}{\delta \G^{\mu\nu}}+D^{\mu}\Big[L D_{\mu}\phi D_{\nu}\phi (1+D_{\a}\phi D^{\a}\phi)+L\G_{\mu\nu}D_{\a}\phi D^{\a}\phi\Big]=0.
\label{45}
\een

\section{Modification of the Friedmann equations}
We consider the gravitational metric $g_{\mu\nu}$ to be a flat Friedmann-Lema{\^i}tre-Robertson-Walker (FLRW) metric and the line element for this is, 
\ben
ds^{2}=dt^{2}-a^{2}(t)\sum_{i=1}^{3} (dx^{i})^{2}, \label{46}
\een
with $a(t)$ being the scale factor as usual. 

From Eq. (\ref{23}), we have the components of the emergent gravity metric as
\ben
&&\G_{00}=(1-\dot\phi^{2})~;~\G_{ii}=-[a^{2}(t)+(\phi^{'})^{2}]~;~\G_{0i}=-\dot\phi \phi^{'}=\G_{i0}, \label{47}
\een
where we consider $\phi\equiv \phi(t,x^{i})$, $\dot\phi=\frac{\p \phi}{\p t}$ and $\phi^{'}=\frac{\p \phi}{\p x^{i}}$. 

So the line element of the FLRW emergent gravity metric is
\ben
dS^{2}=(1-\dot\phi^{2})dt^{2}-\big[a^{2}(t)+(\phi^{'})^{2}\big]\sum_{i=1}^{3} (dx^{i})^{2}-2\dot\phi \phi^{'}dt dx^{i}.
\label{48}
\een

Now from the emergent gravity equation of motion (\ref{21}) we have
\ben
&&\G^{00}(\p_{0}\p_{0}\phi-\ga^{0}_{00}\p_{0}\phi-\ga^{i}_{00}\p_{i}\phi)+\G^{ii}(\p_{i}\p_{i}\phi-\ga^{0}_{ii}\p_{0}\phi-\ga^{i}_{ii}\p_{i}\phi)+\G^{0i}(\p_{0}\p_{i}\phi-\ga^{0}_{0i}\p_{0}\phi-\ga^{i}_{oi}\p_{i}\phi)
\nonumber\\&&+\G^{i0}(\p_{i}\p_{0}\phi-\ga^{0}_{i0}\p_{0}\phi-\ga^{i}_{i0}\p_{i}\phi)=0.
\label{49}
\een

For simplification, we consider the homogeneous {\bf K}-essence scalar field $\phi$, i.e., $\phi(t,x^{i})\equiv\phi(t)$ then $\G_{00}=(1-\dot\phi^{2})$, $\G_{0i}=\G_{i0}=0=\p_{i}\phi$, $\G_{ii}=-a^{2}(t)$ and $X={1\over 2}g^{\mu\nu}\nabla_{\mu}\phi\nabla_{\nu}\phi=\frac{1}{2}\dot\phi^{2}$. This consideration is possible in this case, since the dynamical solutions of the {\bf K}-essence scalar fields spontaneously break Lorentz symmetry. Therefore, the flat FLRW emergent gravity line element (\ref{48}) and the equation of motion (\ref{49}) become
\ben
dS^{2}=(1-\dot\phi^{2})dt^{2}-a^{2}(t)\sum_{i=1}^{3} (dx^{i})^{2},
\label{50}
\een
and
\ben
\frac{\dot a}{a}=H(t)=-\frac{\ddot\phi}{\dot\phi(1-\dot\phi^{2})},
\label{51}
\een
where $H(t)=\frac{\dot a}{a}$ is the usual Hubble parameter (always $\dot a\neq 0$). The Eq. (\ref{51}) gives the relation between Hubble parameter and the time derivatives of the {\bf K}-essence scalar field. Note that, in the above space-time (\ref{50}) always $\dot\phi^{2}<1$. If $\dot\phi^{2}>1$, the signature of this space-time will be ill-defined. Also $\dot\phi^{2}\neq 0$ condition holds good instead of 
$\dot\phi^{2}=0$, which leads to non-applicability of the {\bf K}-essence theory.  
Also, $\dot\phi^{2}\neq 1$ because $\Omega_{matter} +\Omega_{radiation} +\Omega_{dark energy}= 1$ 
and we can measured $\dot\phi^{2}$ as dark energy density in units of the critical density, i.e., it is nothing but $\Omega_{dark energy}$~\cite{gm1,gm2,gm3,gm4}. Therefore $\dot\phi^{2}$ takes the value in between $0$ and $1$.

The Ricci tensors and Ricci scalar of the emergent gravity space-time are 
\ben
\R_{ii}&&=-\frac{a^{2}}{1-\dot\phi^{2}}\left[\frac{\ddot a}{a}+2\left(\frac{\dot a}{a}\right)^{2}+\frac{\dot a}{a}\frac{\dot\phi \ddot\phi}{1-\dot\phi^{2}}\right]=-\frac{a^{2}}{1-\dot\phi^{2}}\left[\frac{\ddot a}{a}+\left(\frac{\dot a}{a}\right)^{2}(2-\dot\phi^{2})\right]\nonumber\\&&=-\frac{a^{2}}{1-\dot\phi^{2}}\left[\dot H+H^{2}(3-\dot\phi^{2})\right],
\label{52}
\een

\ben
\R_{00}=3\frac{\ddot a}{a}+3\frac{\dot a}{a}\frac{\dot\phi \ddot\phi}{1-\dot\phi^{2}}
=3\frac{\ddot a}{a}+3\left(\frac{\dot a}{a}\right)^{2}\dot\phi^{2}=3\left[\dot H+H^{2}(1-\dot\phi^{2})\right],
\label{53}
\een
and
\ben
\R &&= \frac{6}{1-\dot\phi^{2}}\left[\frac{\ddot a}{a}+\left(\frac{\dot a}{a}\right)^{2}+\frac{\dot a}{a}\frac{\dot\phi \ddot\phi}{1-\dot\phi^{2}}\right]= \frac{6}{1-\dot\phi^{2}}\left[\frac{\ddot a}{a}+\left(\frac{\dot a}{a}\right)^{2}(1-\dot\phi^{2})\right]\nonumber\\
&&= \frac{6}{1-\dot\phi^{2}}\left[\dot H +H^{2}(2-\dot\phi^{2})\right],
\label{54}
\een
where we have used the relation (\ref{51}) and $\dot H\equiv\frac{\p H}{\p t}=\frac{a\ddot{a}-\dot{a}^{2}}{a^{2}}$. 

Combining Eqs. (\ref{52}) and (\ref{53}) with (\ref{54})  we get
\ben
\R_{00}=\frac{1}{2}(1-\dot\phi^{2})\R-3H^{2},
\label{55}
\een

\ben
\R_{ii}=-\frac{a^{2}}{(1-\dot\phi^{2})}\left[\frac{1}{6}\R(1-\dot\phi^{2})+H^{2}\right].
\label{56}
\een

We assume that the energy-momentum tensor is an ideal fluid type which is
\ben
T_{\mu}^{\nu}&=&diag(\rho,-p,-p,-p)=(\rho +p)u_{\mu}u^{\nu}-\delta_{\mu}^{\nu} p\nonumber\\
T_{\mu\nu}&=&\G_{\mu\a}T^{\a}_{\nu},
\label{57}
\een
where $p$ is pressure and $\rho$ is the matter density of the cosmic fluid. In the co-moving frame we have $u^{0}=1$ and
$u^{\a}=0$ ; $\a= 1, 2, 3$  in the {\bf K}-essence emergent gravity space-time. 

Now the question is whether this type of energy-momentum tensor is valid or not in case of a perfect fluid model when the kinetic energy ($\dot\phi^{2}$) of the {\bf K-}essence scalar filed is present. The answer is yes since our Lagrangian is $L(X) = 1-\sqrt{1-2X}$. This class of models is equivalent to perfect fluid models with zero vorticity and the pressure (Lagrangian) can be expressed through the energy density only~\cite{babi4,gm4}.

Now we evaluate the $ii$ and $00$ components of the modified field equation (\ref{40}) using (\ref{57}) and considering $\phi\equiv\phi(t)$ only:
\ben
F\R_{ii} + (\bar{G}_{ii}\sq - D_{i}D_{i})F - \frac{1}{2}[f-Lf_{L}(1+\dot\phi^{2})]\bar{G}_{ii}
=\frac{1}{2}f_{L}a^{2}(t)\bar{p}~~~~
\label{58}
\een
and
\ben
F\R_{00} + (\bar{G}_{00}\sq - D_{0}D_{0})F - \frac{1}{2}[f-Lf_{L}(1+\dot\phi^{2})]\G_{00} +\frac{1}{2}Lf_{L}\dot\phi^{2}(1+\dot\phi^{2})
=\frac{1}{2}f_{L}(1-\dot\phi^{2})\bar{\rho},
\label{59}
\een
with $F=f_{\R}(\R,L)\equiv\frac{\p f(\R,L)}{\p \R}$, $\bar{p}=p(1+\dot\phi^{2})$ and $\bar{\rho}=\rho(1+\dot\phi^{2})$. 

Now, we calculate the terms $\G_{00}\sq F$ and $\G_{ii}\sq F$, using the determinant of the flat FLRW emergent gravity metric $\s{-\G}=a^{3}\s{1-\dot\phi^{2}}$ and the relation (\ref{51}):
\ben
\G_{00}\sq F
=\ddot F+3 \frac{\dot a}{a}\dot F +\dot F\frac{\dot\phi \ddot\phi}{(1-\dot\phi^{2})}=\ddot F + H\dot F (3-\dot\phi^{2}),~~~~
\label{60}
\een
and
\ben
\G_{ii}\sq F
=D_{i}D_{i}F-\frac{a^{2}}{(1-\dot\phi^{2})}\left[\ddot F +2H\dot{F}(1-\dot\phi^{2})\right],
\label{61}
\een
where we have used, $(\p_{i}t)^{2}=\frac{a^{2}}{1-\dot\phi^{2}}$ for flat FLRW emergent gravity metric.

Now we substitute Eqs. (\ref{55}) and  (\ref{60}) in the above Eq. (\ref{59}), so that we have the first modified Friedmann equation as
\ben
3H^{2}&&=\frac{1}{F}[-\frac{1}{2}\bar{\rho}f_{L}(1-\dot\phi^{2})+3H\dot{F}+(1-\dot\phi^{2})\frac{1}{2}(F\R-f)+\frac{1}{2}Lf_{L}(1+\dot\phi^{2})]\nonumber\\&&=\frac{1}{F}\Big[-\frac{1}{2}\bar{\rho}f_{L}(1-\dot\phi^{2})
+3H\dot{\R}F_{\R}
+3HF_{L}L_{X}\dot{X}+(1-\dot\phi^{2})\frac{1}{2}(F\R-f)+\frac{1}{2}Lf_{L}(1+\dot\phi^{2})\Big].~~~\label{62}
\een

We also substitute Eqs. (\ref{52}), (\ref{54}) and (\ref{61}) in the $ii$-components of Eq. (\ref{58}) and thereafter rearranging, we get the second modified Friedmann equation for the flat FLRW {\bf K}-essence emergent gravity space-time under $f(\R, L(X))$ theory. Hence
\ben
&&2\dot{H}+H^{2}(3-2\dot\phi^{2})
=\frac{1}{F}\Big[\frac{1}{2}\bar{p}f_{L}(1-\dot\phi^{2})+\ddot{F}+2H\dot{F}(1-\dot\phi^{2})-\frac{1}{2}(1-\dot\phi^{2})(f-\R F)+\frac{1}{2}Lf_{L}(1-\dot\phi^{2})(1+\dot\phi^{2})\Big]\nonumber\\&&=\frac{1}{F}\Big[\frac{1}{2}\bar{p}f_{L}(1-\dot\phi^{2})+\ddot{\R}F_{\R}+(\dot{\R})^{2}F_{\R\R}+2H\dot{\R}F_{\R}(1-\dot\phi^{2})-\frac{1}{2}(1-\dot\phi^{2})(f-\R F)+\frac{1}{2}Lf_{L}(1-\dot\phi^{2})(1+\dot\phi^{2})\Big]\nonumber\\&&+\frac{1}{F}\Big[2H(1-\dot\phi^{2})F_{L}L_{X}\dot{X}+F_{LL}(L_{X}\dot{X})^{2}+F_{L}L_{XX}(\dot{X})^{2}+F_{L}L_{X}\ddot{X}\Big]. 
\label{63}
\een

Since the  Lagrangian ($L$) of the {\bf K}-essence theory is function of $X(={1\over 2}g^{\mu\nu}\nabla_{\mu}\phi\nabla_{\nu}\phi)$, therefore we can write
\ben
&&\dot{F}=F_{\R}\dot{\R}+F_{L}L_{X}\dot{X}~and~ \nonumber\\&&\ddot{F}=F_{\R}\ddot{\R}+(\dot{\R})^{2}F_{\R\R}+F_{LL}(L_{X}\dot{X})^{2}+F_{L}L_{XX}(\dot{X})^{2}+F_{L}L_{X}\ddot{X}.
\label{64}
\een

The above Friedmann equations in the presence of the kinetic energy of the {\bf K}-essence scalar field are different from the usual $f(R)$ gravity model. Note that, if we consider $f(\R,L(X))\equiv f(R)$ and $\G_{\mu\nu}\equiv g_{\mu\nu}$ then, the above modified Friedmann equations (\ref{62}) and (\ref{63}) reduces to the usual Friedmann equations of $f(R)$ gravity, with $\kappa=1$ and $T_{\mu\nu}$ replaced by $\frac{1}{2}T_{\mu\nu}$~\cite{sotiriou,felice} as
\ben
3H^{2}=\frac{1}{F}[\frac{-\rho}{2}+\frac{RF-f}{2}-3H\dot{R}F_{R}],
\label{65}
\een
and
\ben
2\dot{H}+3H^{2}=\frac{1}{F}\Big[\frac{p}{2}+(\dot{R})^{2}F_{RR}+2H\dot{R}F_{R}+\ddot{R}F_{R}-\frac{f-RF}{2}\Big].
\label{66}
\een

\section{Solution to the Friedmann equation using power law}
We choose Starobinsky type model~\cite{staro,sotiriou} to investigate our theory and evaluate some cosmological values of the universe. This model has achieved popularity as, the inflationary predictions produced by this theory seems very much consistent with the observational data. The coefficient of $R^2$ curvature term single-handedly obtains the slow-roll inflation with tremendous success, without the introduction of outside inflation field by hand. Other reasons for Starobinksy model to be treated as an important model has been discussed in~\cite{Kehagias}.
Now we write
$f(\R,L)$ as 
\ben f(\R,L)=\R+\alpha\R^2+L,
\label{67}
\een
and $L$ is the DBI type Lagrangian mentioned in (9).
Therefore, we get
\ben
f_L=\frac{\partial f}{\partial L}=1,
\quad{F}=\frac{\partial f}{\partial \R}=1+2\alpha\R, \quad{F}_{\R}=2\alpha, \quad{F}_L=0.
\label{69}
\een

Using these values and after some algebraic calculations we can write Friedmann equation (\ref{62}) as
\ben
(1+2\alpha \R)3H^{2}=-\frac{1}{2}\bar{\rho}(1-\dot{\phi}^2)+\frac{1}{2}(1-\dot{\phi}^2)\alpha \R^{2}+6\alpha H\dot{\R}+\dot{\phi}^2(1-\sqrt{1-\dot{\phi}^2}).
\label{70}
\een

Analogous to~\cite{Goheer1,Goheer2,Singh}, let us now assume, there exists an exact power–law solution to the field equations, i.e., the
scale factor behaves as
\ben
a(t)=a_0 t^m, \label{71}
\een
where $m(>0)$ is a fixed real number.

The definition of $H$ gives us
\ben
H=\frac{\dot{a}}{a}=\frac{m}{t}, \quad
\dot{H}=-\frac{m}{t^2},
\ddot{H}=\frac{2m}{t^3}, \quad
\dot{\ddot{H}}=-\frac{6m}{t^4}.
\label{73}
\een

Now, taking Eq. (\ref{54}) into consideration we can evaluate the value of Ricci scalar as
\ben
\R=\frac{6}{(1-\dot{\phi}^2)t^2}\Big[-m+m^2 (2-\dot{\phi}^2)\Big],
\label{74}
\een
and then using Eq. (\ref{51}) we get
\ben
\dot{\R}=\frac{6}{1-\dot\phi^{2}}\Big[\ddot{H}+4H\dot{H}(1-\dot\phi^{2})-2H^{3}\dot\phi^{2}\Big]=\frac{12}{t^3 (1-\dot{\phi}^2)}\Big[m-m^3\dot{\phi}^2 -2m^2 (1-\dot{\phi}^2)\Big].~~~~
\label{75}
\een

Now putting the values of Eqs. (\ref{71}) -- (\ref{75}) in (\ref{70}), we simply get
\ben
\frac{1}{2}\bar{\rho}(1-\dot\phi^{2})&&=\frac{90\alpha m^2}{t^4 (1-\dot{\phi}^2)}-\frac{180\alpha m^3}{t^4(1-\dot{\phi}^2)}+\frac{180\alpha m^3\dot{\phi}^2}{t^4(1-\dot{\phi}^2)}\nonumber\\
&&-\frac{108\alpha m^4 \dot{\phi}^2}{t^4 (1-\dot{\phi}^2)}+\frac{18\alpha m^4 \dot{\phi}^4}{t^4 (1-\dot{\phi}^2)}-\frac{3m^2}{t^2}+\dot{\phi}^2(1-\sqrt{1-\dot{\phi}^2}).
\label{76}
\een

Now from the second Friedmann equation (\ref{63}), we have
\ben
(1+2\alpha \R)\Big[2\dot{H}+H^2(3-2\dot{\phi}^2)\Big]=\frac{1}{2}\bar{p}(1-\dot{\phi}^2)+2\alpha \ddot{\R}\nonumber\\
+4\alpha H \dot{\R}(1-\dot{\phi}^2)+\frac{1}{2}\alpha \R^2(1-\dot{\phi}^2)
+\dot{\phi}^2(1-\sqrt{1-\dot{\phi}^2})~~~~~
\label{77}
\een
or
\ben
\frac{1}{2}\bar{p}(1-\dot{\phi}^2)&&=-\frac{2m}{t^2}+\frac{3m^2}{t^2}+\frac{2m\dot{\phi}^2}{t^2}-\frac{5m^{2}\dot\phi^{2}}{t^{2}}-\frac{186\alpha m^2}{t^4(1-\dot{\phi}^2)}
+\frac{240\alpha m^2\dot{\phi}^2}{t^4(1-\dot{\phi}^2)}
+\frac{84\alpha m^3}{t^4(1-\dot{\phi}^2)}-\frac{252\alpha m^3\dot{\phi}^2}{t^4(1-\dot{\phi}^2)}
+\frac{12\alpha m^4 \dot{\phi}^2}{t^4(1-\dot{\phi}^2)}\nonumber\\&&+\frac{72\alpha m}{t^4(1-\dot{\phi}^2)}-\frac{42\alpha m^4 \dot{\phi}^4}{t^4(1-\dot{\phi}^4)}
+\frac{96\alpha m^3 \dot{\phi}^4}{t^4(1-\dot{\phi}^2)}-\frac{1}{2}\dot{\phi}^2(1-\sqrt{1-\dot{\phi}^2})+\frac{1}{2}\dot{\phi}^4(1-\sqrt{1-\dot{\phi}^2}).~~~
\label{78}
\een

For our case, the energy-momentum conservation relation is
\ben
D^{\mu}\bar{T}_{\mu\nu}=0,
\label{79}
\een
with $\T_{\mu\nu}=T_{\mu\nu}[1+D_{\a}\phi D^{\a}\phi]$. 

Now, using Eqs. (\ref{57}) and (\ref{79}), we have the conserving equation as
\ben
\dot{\bar{\rho}}=3\frac{\dot{a}}{a}(\bar{\rho}+\bar{p}), 
\label{80}
\een
where $\bar{\rho}$ and $\bar{p}$ already have been defined. It is essential to mention here that, $\bar{\rho}$ and $\bar{p}$ are not same as the normal $\rho$ and $p$. 

Now considering the power law, we get from Eq. (\ref{80})
\ben
\bar{\rho}=\rho(1+\dot{\phi}^2)=\rho_0 t^{-3m(1+\omega)},
\label{81}
\een
where $\omega=\frac{\bar{p}}{\bar{\rho}}=\frac{p}{\rho}$.

Now, putting the value of $\bar{\rho}$ in Friedmann equation (\ref{76}) we have
\ben
\frac{1}{2}\rho_0 t^{-3m(1+\omega)}&&=\frac{90\alpha m^2}{t^4 (1-\dot{\phi}^2)^{2}}-\frac{180\alpha m^3}{t^4(1-\dot{\phi}^2)^{2}}+\frac{180\alpha m^3\dot{\phi}^2}{t^4(1-\dot{\phi}^2)^{2}}
-\frac{108\alpha m^4 \dot{\phi}^2}{t^4 (1-\dot{\phi}^2)^{2}}+\frac{18\alpha m^4 \dot{\phi}^4}{t^4 (1-\dot{\phi}^2)^{2}}\nonumber\\&&-\frac{3m^2}{t^2(1-\dot{\phi}^2)}+\frac{\dot{\phi}^2}{(1-\dot{\phi}^2)}(1-\sqrt{1-\dot{\phi}^2}).~~~
\label{82}
\een

On the other hand, to maintain the energy-momentum conservation, the relation (\ref{45}) must be satisfied. So, the effective functional relation (\ref{45}) for homogeneous {\bf K}-essence scalar field reduces to
\ben
3\dot\phi^{2}-2=2\sqrt{1-\dot\phi^{2}},
\label{83}
\een
where we have used Eqs. (\ref{22}) and (\ref{69}).

Solving the above Eq. (\ref{83}) we have either, $\dot\phi^{2}=0$, which is not acceptable for our case or
\ben
\dot\phi^{2}=\frac{8}{9}=0.888=\text{constant}.
\label{84}
\een

It should be noted that, the exact solution of field equations (Eq. no. 2 in~\cite{Singh}) are already obtained by the assumption of power law form of the scale factor using Starobinsky Model in~\cite{Singh}. The results of that case is
\ben
\rho_{\phi}=\frac{3n^2}{t^2}-\frac{\rho_{0}}{t^{3n(1+\omega)}}+\frac{54\alpha n^2(2n-1)}{t^4},
\label{85}
\een
and
\ben
p_{\phi}=\frac{n(2-3n)}{t^{2}}+\frac{18\a n(2n-1)(4-3n)}{t^{4}}-\frac{\o\rho_{m0}}{t^{3n(1+\o)}},
\label{86}
\een
where $\rho_{\phi}$ and $p_{\phi}$ is the energy density and pressure of the scalar field and $n$ is synonymous to $m$ for our case. 

Rearranging Eq. (\ref{85}) we get
\ben
\rho_0 t^{-3n(1+\omega)}=\frac{3n^2}{t^2}+\frac{108\alpha n^3}{t^4}-\frac{54\alpha n^2}{t^4} 
-\frac{1}{2}\dot{\phi}^2-V(\phi),
\label{87}
\een
where they have defined, $\rho_{\phi}=\frac{1}{2}\dot{\phi}^2+V(\phi)$, $p_{\phi}=\frac{1}{2}\dot{\phi}^2-V(\phi)$, $V(\phi)$ is the scalar potential. 

Singh et al.~\cite{Singh} in their paper have used canonical Lagrangian and usual field equations of $f(R)$-gravity but, in our case we have used non-canonical Lagrangian and corresponding field equations (\ref{40}). This is the basic difference between these two articles. Also, it should be clearly mentioned that, the scalar field of them is not identical with the {\bf K}-essence scalar field.

Now let us concentrate upon the deceleration parameter using the expression
\ben
q=-\frac{1}{H^2}\frac{\ddot{a}}{a}=\frac{1}{m}-1,
\label{dec}
\een
where we have used Eq. (\ref{71}).
From the above expression, it is obvious that for our present epoch the deceleration parameter should have a negative value to support the acceleration of the universe. Therefore, we can conclude from Eq. (\ref{dec}) that the $m$ takes the value greater than $1$. Negative value of $m$ can not be considered, since observations show  the universe is expanding.

As we know that, the value of $\dot{\phi}^2$ is less than $1$, so neglecting the higher order terms $O(\dot{\phi}^4$) in Eq. (\ref{76}) and (\ref{78}), and using Eq. (\ref{84}), we get the equation of state (EOS) parameter of this scenario as:
\ben
\omega=\frac{-(2+13m)t^{2}+\alpha (648+246m-1260m^{2}-96m^{3})}{-3mt^{2}+\alpha m(810-180m-864m^{2})}.\nonumber\\
\label{eos1}
\een

\begin{figure}[h]
\centering
\includegraphics[width=10cm, height=10cm]{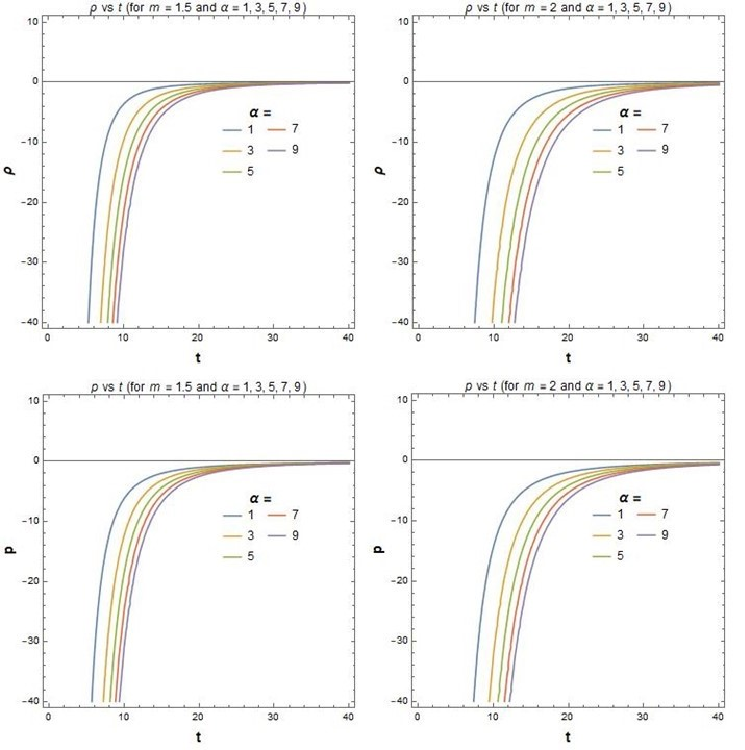}
 \caption{(Color online) Variation of $\rho$ and $p$ with $t$ for different values of $m$ ($=1.5,2$) and $\a$$(=1,3,5,7,9)$}\label{Fig1}
\end{figure}

\begin{figure}[h]
\centering
\includegraphics[width=10cm, height=10cm]{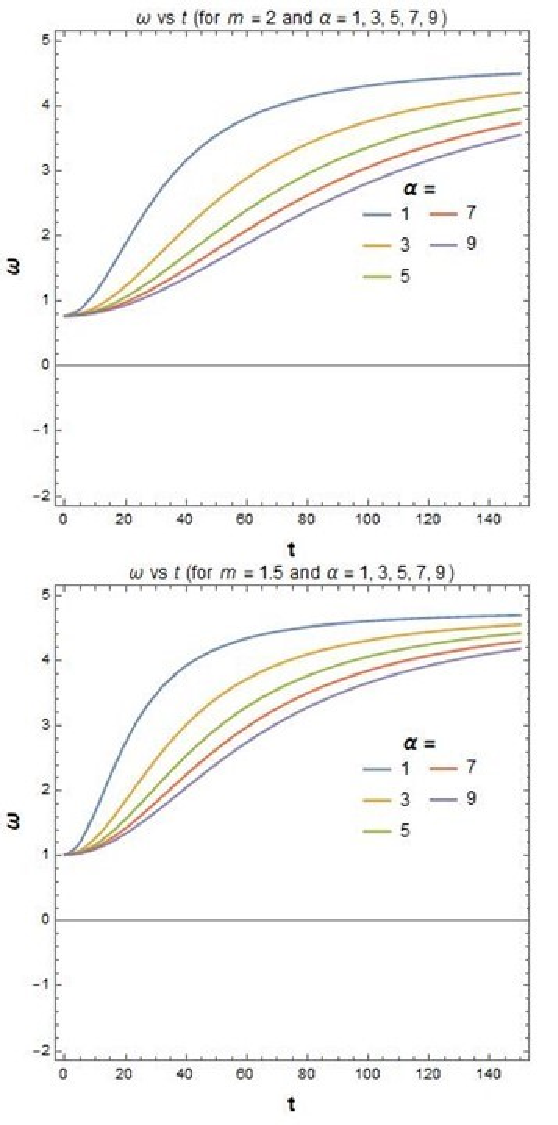}
 \caption{(Color online) Variation of $\omega$ with $t$ for different values of $m$ ($=1.5,2$) and $\a$$(=1,3,5,7,9)$}\label{Fig2}
\end{figure}

The variation of $\rho$ and $p$ (using Eqs. (\ref{76}) and (\ref{78}) and omitting $O(\dot{\phi}^4$) terms) with time ($t$) has been plotted in Fig. 1 for different choices of the positive power law parameter ($m=1.5,2$) and the positive coefficient of $\R^2$ in Starobinsky model ($\alpha=1,3,5,7,9$). Fig. 2 shows the variation of EOS parameter $\omega$ with $t$ for the aforementioned values of $m$ and $\alpha$. As we know that, the values of $\rho$ and $p$ should differ in signature for dark energy dominated era and simultaneously the value of EOS parameter ($\omega$) should approach a value close to $-1$, therefore, the above two figures concludes that the choice of {\it positive $m$ and positive $\alpha$ is ruled out} for our model to produce dark energy conditions. The time $t$ here is the cosmological time i.e., the time corresponding to the FLRW metric.

The negative values of $\alpha$ has already been considered in~\cite{Moraes} for $f(R,T)$ gravity. So, let's check the results of our model for positive value of $m(=2,3)$ and negative values of $\alpha(=-0.9,-0.7,-0.5,-0.3,-0.1)$. Fig. 3 depicts the variation of $\rho$ and $p$ with time ($t$) for above values of the parameters. Fig. 4 shows the variation of EOS parameter $(\omega)$ with time $(t)$. From Fig. 3, it is evident that the value of $\rho$ and $p$ have the expected nature at a particular region of time. Simultaneously, Fig. 4 produces the anticipated value of $\omega$ which is $-1$ for dark energy epoch.  We discuss the results more elaborately obtained in Figs. 3 and 4 in the following subsection.

\begin{figure}[h]
\centering
\includegraphics[width=10cm, height=11cm]{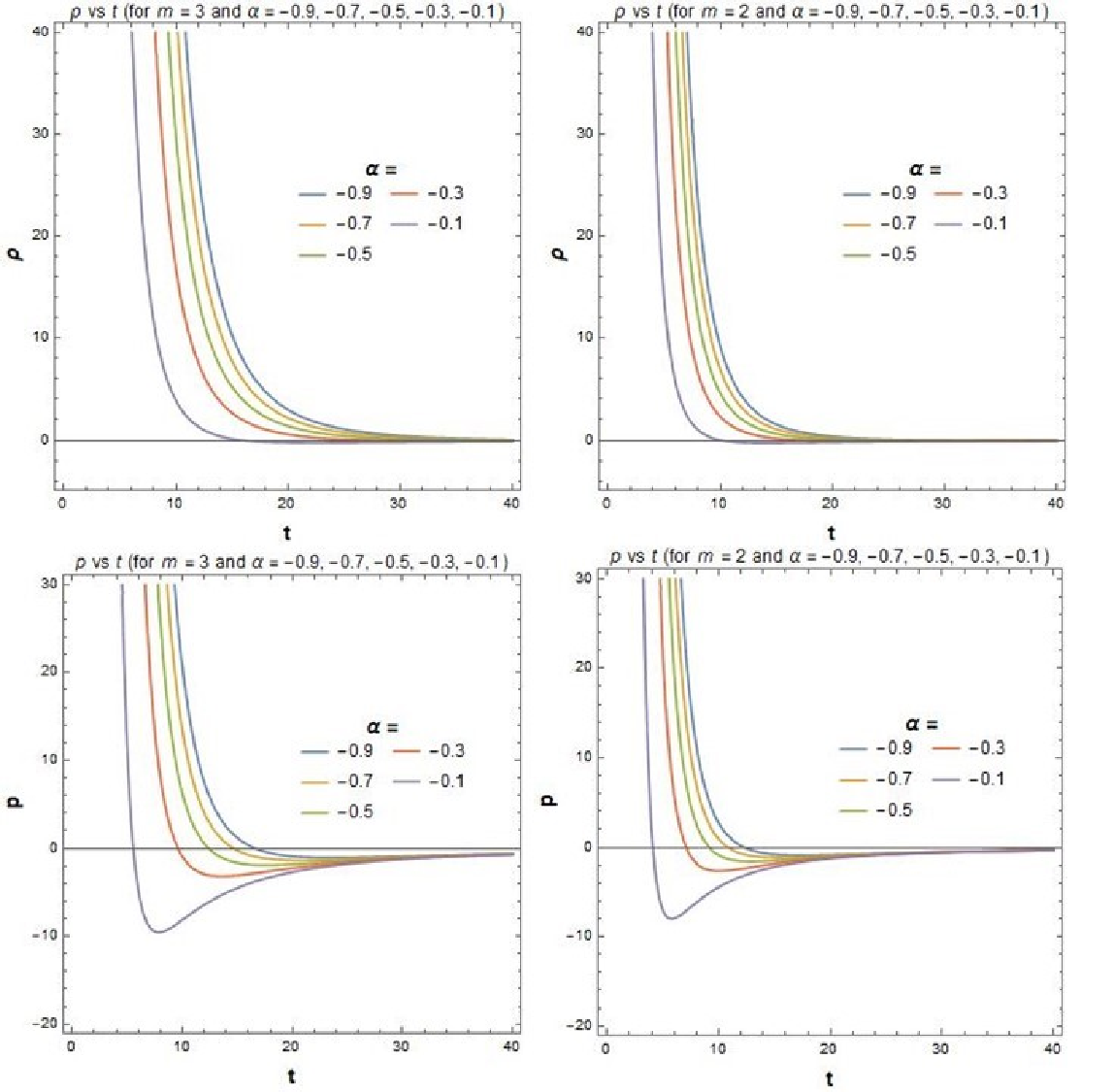}
 \caption{(Color online) Variation of $\rho$ and $p$ with $t$ for different values of $m$ ($=3,2$) and $\a(=-0.9,-0.7,-0.5,-0.3,-0.1)$}\label{Fig3}
\end{figure}

\begin{figure}[h]
\centering
\includegraphics[width=10cm, height=7cm]{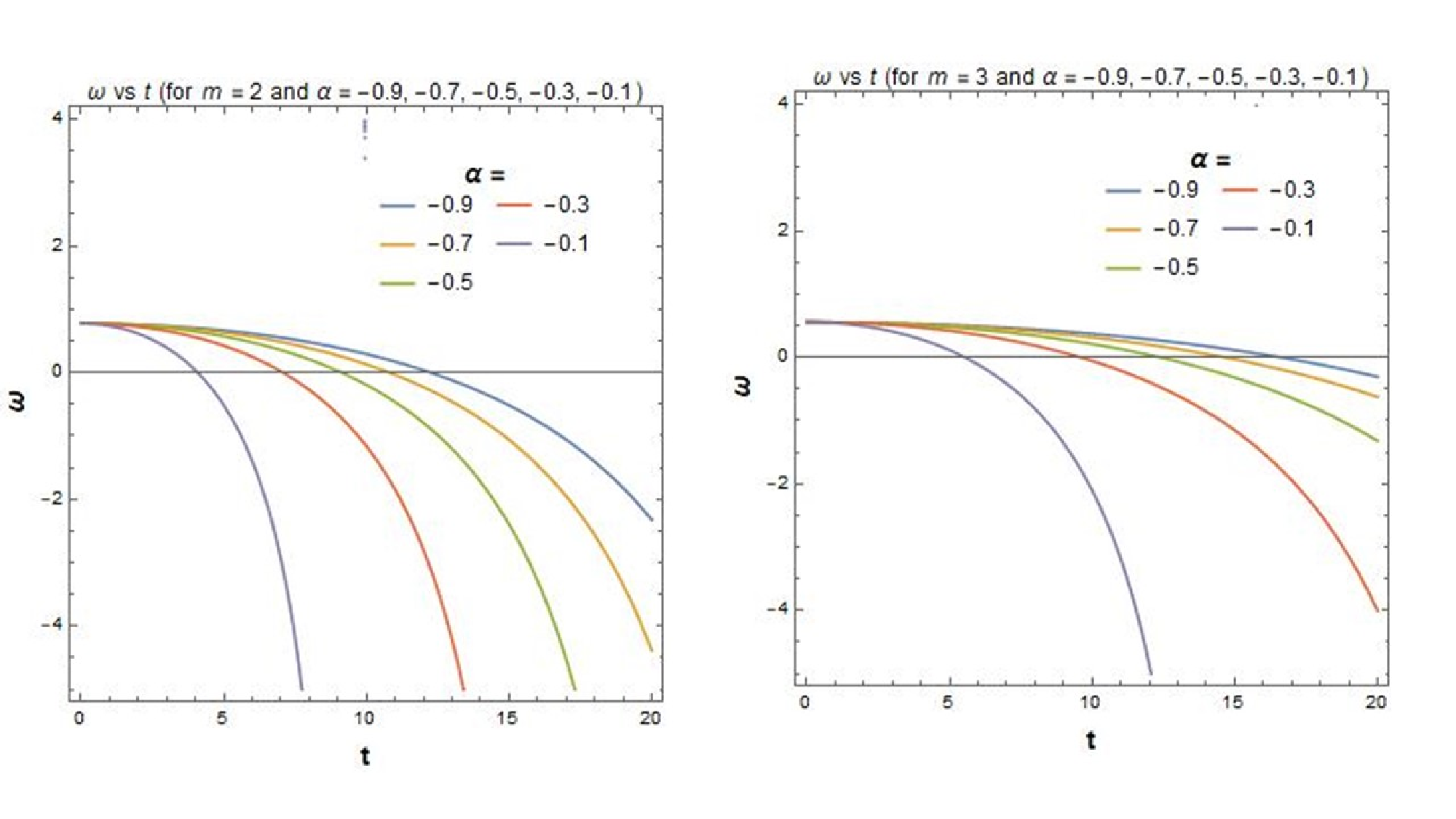}
 \caption{(Color online) Variation of $\omega$ with $t$ for different values of $m$ ($=2, 3$) and $\a$$(=-0.1,-0.3,-0.5,-0.7,-0.9)$}\label{Fig4}
\end{figure}

\subsection{Observational Verification of the Model}
Before entering into this section we would like to discuss two significant works, one has been done by Tripathi et al.~\cite{Tripathi} and the other one is done by Moraes et al.~\cite{Moraes}. In~\cite{Tripathi} they have constrained the dark energy models for low redshifts and compared the data with the observations of  Supernova Type Ia data, Baryon Acoustic Oscillation data and Hubble parameter measurements. On the other hand, in~\cite{Moraes} the authors have studied various cosmological aspects with the help of Starobinsky model in the framework of $f(R,T)$ gravity. They have found the nature of material content of the universe, i.e. $\rho$ and $p$ in both decelerated and accelerated regime of the universe.


\begin{table}[h!]
\begin{center}
\resizebox{8.5cm}{!}{
\begin{tabular}{ |c|c|c|c|c|c| } 
\hline
$m$ & $\alpha$ & $t$ & $\omega$ ($3\sigma$ confidence)  & Observation\\
\hline
 2 & \multirow{2}{4em}{$-0.9$} & $14.96-15.5$ & \multirow{10}{4em}{$-0.95\geq\omega\geq-1.13$}  & \multirow{10}{4em}{SNIa+ BAO+ H(z)} \\ 
 3 &  & $22.8-23.87$ &  \\
 2 & \multirow{2}{4em}{$-0.7$} & $13.2-13.67$ & \\ 
 3 &  & $20.1-21.05$ & \\
 2 & \multirow{2}{4em}{$-0.5$} & $11.16-11.55$ & \\
 3 &  & $17.06-17.8$ & \\
 2 & \multirow{2}{4em}{$-0.3$} & $8.64-8.95$ & \\
 3 &  & $13.22-13.78$ & \\
 2 & \multirow{2}{4em}{$-0.1$} & $4.98-5.16$ & \\
 3 &  & $7.63-7.97$ & \\
\hline
\end{tabular}}
\end{center}
\caption{Table for Observational Verification of the Model.}
 \end{table}


The variation of $\rho$ and $p$ with time obtained in Fig. 3 is quite similar with the variation obtained in~\cite{Moraes}, though our models differ from each other. Fig. 3 shows that at early time ($t\rightarrow 0$) the  pressure was positive. But after a certain value of time it takes negative value which may be correlated with the effect of the negative pressure fluid, responsible for the accelerating universe. We have shown a table which depicts that for different choices of the positive $m$ and the negative $\alpha$, we get a range of $t$ (from Fig. 4) where the value of $\omega$ satisfies with the observational data of Supernova Type Ia data, Baryon Acoustic Oscillation data and Hubble parameter  measurements (Observational data are taken from~\cite{Tripathi}). Furthermore, if we concentrate on the range of $t$ that has been shown in Table 1 and match those values with Fig. 3, then can be observed that at those particular region of time, the value of $p$ takes the negative sign whereas the $\rho$ is positive.

\section{Energy conditions in $f(\R, L(X))$-Gravity}
With the help of the modified field equation (\ref{40}) for $f(\R,L(X))$ theory, the emergent Einstein's equation (\ref{28}) can be written as ($\k=1$)
\ben
\R_{\mu\nu}-\frac{1}{2}\G_{\mu\nu}\R=T_{\mu\nu}^{eff},
\label{89}
\een
where
\ben
T_{\mu\nu}^{eff}&&=\frac{1}{F}[\frac{1}{2}f_{L}\T_{\mu\nu}-\frac{1}{2}\R F\G_{\mu\nu}-(\G_{\mu\nu}\bar{\square}-D_{\mu}D_{\nu})F+\frac{1}{2}\G_{\mu\nu}(f-Lf_{L}(1+D_{\a}\phi D^{\a}\phi))-\frac{1}{2}Lf_{L}D_{\mu}\phi D_{\nu}\phi(1+D_{\a}\phi D^{\a}\phi)]
\nonumber\\&&
=\frac{1}{F}[\frac{1}{2}f_{L}\T_{\mu\nu}+\frac{1}{2}\G_{\mu\nu}(f-F\R)-(\G_{\mu\nu}\bar{\square}-D_{\mu}D_{\nu})F-\frac{1}{2}Lf_{L}\G_{\mu\nu}(1+D_{\mu}\phi D^{\mu}\phi)^{2}],
\label{90}
\een
with $F=f_{\R}=\frac{\p f(\R,L(X))}{\p \R}$.

The trace of the effective energy momentum tensor (\ref{90}) is
\ben
T^{eff}=\frac{1}{F}[\frac{1}{2}f_{L}\T +2(f-F \R)-3 \bar{\square} F-2Lf_{L}(1+D_{\mu}\phi D^{\mu}\phi)^{2}].
\label{91}
\een

Now from Eq. (\ref{89}), we have the emergent Ricci tensor in terms of the effective energy momentum tensor as
\ben
\R_{\mu\nu}=T_{\mu\nu}^{eff}-\frac{1}{2}\G_{\mu\nu}T^{eff}.
\label{92}
\een
   
Let $\bar{u}^{\mu}$ be the tangent vector field to a congruence of time-like geodesics in the {\bf K}-essence emergent space-time manifold endowed with the metric $\G_{\mu\nu}$ ($\G_{\mu\nu}\bar{u}^{\mu}\bar{u}^{\nu}=1$), then the strong energy condition (SEC) (\ref{9}) in $f(\R,L(X))$ modified gravity can be expressed as
\ben
\R_{\mu\nu}\bar{u}^{\mu}\bar{u}^{\nu}=(T_{\mu\nu}^{eff}\bar{u}^{\mu}\bar{u}^{\nu}-\frac{1}{2}T^{eff})\geq 0.
\label{93}
\een

On the other hand, if we consider $\bar{k}^{\mu}$ be the tangent vector along the null geodesic congruence ($\G_{\mu\nu}\bar{k}^{\mu}\bar{k}^{\nu}=0$), then the null energy condition (NEC) (\ref{11}) in $f(\R,L(X))$  gravity is
\ben
\R_{\mu\nu}\bar{k}^{\mu}\bar{k}^{\nu}=T_{\mu\nu}^{eff}\bar{k}^{\mu}\bar{k}^{\nu}\geq 0.
\label{94}
\een

So, considering an additional condition~\cite{wang2} $\frac{f_{L}(\R,L)}{f_{\R}(\R,L)}>0$, and the {\bf K}-essence scalar field to be homogeneous, i.e., $\phi(x^{i},t)\equiv \phi(t)$ and using the perfect fluid energy momentum tensor (\ref{57}), we have the SEC and NEC in the $f(\R,L(X))$ gravity are
\ben
SEC&:&~\bar{\rho}+3\bar{p}-\frac{2}{f_{L}}(f-F\R)+2L(1+\dot\phi^{2})^{2}+\frac{6}{f_{L}(1-\dot\phi^{2})}[\ddot{F}(1-\frac{2}{3}\dot\phi^{2})+H\dot{F}(1-\dot\phi^{2}+\frac{2}{3}\dot\phi^{4})]\geq 0,
\label{95}\\
NEC&:&~~\bar{\rho}+\bar{p}+\frac{2}{f_{L}}(\ddot{F}-H\dot{F}\dot\phi^{2})\geq 0,
\label{96}
\een
where $\bar{\rho}=\rho(1+\dot\phi^{2})$ and $\bar{p}=p(1+\dot\phi^{2})$.

To evaluate the effective density $\bar{\rho}^{eff}$ and effective pressure $\bar{p}^{eff}$ in the {\bf K}-essence emergent $f(\R,L(X))$ gravity, we consider two following equations
\ben
T_{\mu\nu}^{eff}\bar{u}^{\mu}\bar{u}^{\nu}-\frac{1}{2}\G_{\mu\nu}T^{eff}\bar{u}^{\mu}\bar{u}^{\nu}=\bar{\rho}^{eff}+3\bar{p}^{eff}
\label{97}
\een
and
\ben
T_{\mu\nu}^{eff}\bar{k}^{\mu}\bar{k}^{\nu}=\bar{\rho}^{eff}+\bar{p}^{eff}.
\label{98}
\een

Solving these Eqs. (\ref{62}) and (\ref{63}), we get
\ben
\bar{\rho}^{eff}&=&\bar{\rho}+\frac{1}{f_{L}}(f-F\R)-\frac{6}{f_{L}(1-\dot\phi^{2})}[\frac{1}{3}\ddot{F}\dot\phi^{2}+H\dot{F}(1-\frac{1}{3}\dot\phi^{4})]-L(1+\dot\phi^{2})^{2},
\label{99}\\
\bar{p}^{eff}&=&\bar{p}-\frac{1}{f_{L}}(f-F\R)+\frac{3}{f_{L}(1-\dot\phi^{2})}[\frac{1}{3}\ddot{F}(2-\dot\phi^{2})+H\dot{F}(1-\frac{2}{3}\dot\phi^{2}+\frac{1}{3}\dot\phi^{4})]+L(1+\dot\phi^{2})^{2}.
\label{100}
\een

From the above Eqs. (\ref{99}) and (\ref{100}), we have WEC and DEC respectively for the {\bf K}-essence emergent\\ $f(\R,L(X))$ gravity as 
\ben
WEC&:&~~\bar{\rho}+\frac{1}{f_{L}}(f-F\R)-\frac{6}{f_{L}(1-\dot\phi^{2})}\Big[\frac{1}{3}\ddot{F}\dot\phi^{2}+H\dot{F}(1-\frac{1}{3}\dot\phi^{4})\Big]-L(1+\dot\phi^{2})^{2}\geq 0,
\label{101}\\
DEC~&:&\bar{\rho}-\bar{p}+\frac{2}{f_{L}}(f-F\R)-\frac{2}{f_{L}(1-\dot\phi^{2})}[\ddot{F}(1+\frac{1}{2}\dot\phi^{2})+3H\dot{F}(\frac{3}{2}-\frac{1}{3}\dot\phi^{2}-\frac{1}{6}\dot\phi^{4})]-2L(1+\dot\phi^{2})^{2}\geq 0.
\label{102}
\een

These energy conditions (\ref{95}), (\ref{96}), (\ref{101}) and (\ref{102}) of the {\bf K}-essence emergent $f(\R,L(X))$ gravity are different from the usual $f(R,L_{m})$-gravity (\ref{13}) and $f(R)$-gravity (\ref{12}) in the presence of the {\bf K-}essence scalar field $\phi$. Also note that if we consider $f(\R,L(X))\equiv R$ and $\G_{\mu\nu}\equiv g_{\mu\nu}$, then we can get back to the usual energy conditions of GR, i.e., $SEC:~\rho+3p\geq 0$; $NEC:~\rho+p\geq 0$; $WEC:~\rho\geq 0$ and $DEC:~\rho\geq |p|$.

One may notice that we briefly have discussed the energy conditions of $f(R)$ and $f(R,L_m)$ gravity in Appendix.

\subsection{Constraints on {\bf K}-essence emergent $f(\R,L(X))$-gravity}
The inequalities of the energy conditions (\ref{95}), (\ref{96}), (\ref{101}) and (\ref{102}) can also be expressed in terms of the deceleration ($q$), jerk ($j$), and snap ($s$) parameters such that the Ricci scalar and its derivatives for a spatially flat {\bf K}-essence emergent FLRW geometry (\ref{50}) are
\ben
\R=\frac{6}{1-\dot\phi^{2}}\left[\dot H +H^{2}(2-\dot\phi^{2})\right]=\frac{6H^{2}}{1-\dot\phi^{2}}\left[1-q-\dot\phi^{2}\right],\label{103}
\een
\ben
\dot{\R}=\frac{6}{1-\dot\phi^{2}}\left[\ddot{H}+4H\dot{H}(1-\dot\phi^{2})-2H^{3}\dot\phi^{2}\right]
=\frac{6H^{3}}{1-\dot\phi^{2}}\left[(j-q-2)+2\dot\phi^{2}(1-2q)\right], \label{104}
\een
\ben
&&\ddot{\R}=\frac{6}{1-\dot\phi^{2}}[(\dot{\ddot{H}}+4\dot{H}^{2}+4H\ddot{H})-2\dot\phi^{2}(2\dot{H}^{2}+3H\ddot{H}-2H^{4}+3H^{2}\dot{H}]
\nonumber\\&&=\frac{6H^{4}}{1-\dot\phi^{2}}[(s+q^{2}+8q+6)-2\dot\phi^{2}(3+3j+10q+2q^{2})], \label{105}
\een
where~\cite{wang2,santos,harr,land,visser1,visser2}
\ben
q=-\frac{1}{H^{2}}\frac{\ddot{a}}{a}~;~j=\frac{1}{H^{3}}\frac{\dot{\ddot{a}}}{a}~;~s=\frac{1}{H^{4}}\frac{\ddot{\ddot{a}}}{a}.
\label{106}
\een

Now from Eq. (\ref{64}), we evaluate the values of $\dot{F}$ and $\ddot{F}$ (using Eq. (\ref{51})) in terms of $q,~ j,~ s$ as
\ben
\dot{F}&=&\frac{6H^{3}}{1-\dot\phi^{2}}F_{\R}[(j-q-2)+2\dot\phi^{2}(1-2q)]-HF_{L}L_{XX}\dot\phi^{2}(1-\dot\phi^{2}),
\label{107}\\
\ddot{F}&=&\frac{6H^{4}}{1-\dot\phi^{2}}F_{\R}[(s+q^{2}+8q+6)-2\dot\phi^{2}(3+3j+10q+2q^{2})]+\frac{36H^{6}}{(1-\dot\phi^{2})^{2}}F_{\R\R}[(j-q-2)+2\dot\phi^{2}(1-2q)]^{2}\nonumber\\&&+H\dot\phi^{4}(1-\dot\phi^{2})^{2}\left(F_{LL}L_{X}^{2}+F_{L}L_{XX}\right)-F_{L}L_{X}\dot\phi^{2}(1-\dot\phi^{2})[\dot{H}-2H^{2}(1-2\dot\phi^{2})].~~
\label{108}
\een

Therefore, putting these values of  $\dot{F}$ and $\ddot{F}$ in the energy conditions (\ref{95}), (\ref{96}), (\ref{101}) and (\ref{102}) we have the energy conditions in terms of $q$,~$j$~and~$s$. We can easily check that these energy conditions in terms of $q$,~$j$~and~$s$ are also different from the $f(R,L_{m})$-gravity~\cite{wang2} in the presence of the {\bf K}-essence scalar field $\phi$.\\

\subsection{An example of the energy conditions}
Considering the Starobinsky Model, i.e., Eq. (\ref{67}) and hence we obtain the following results as
$f_{L}=1$,~$F=1+2\alpha\R$,~$\dot{F}=2\alpha\dot{\R}$~and~$\ddot{F}=2\alpha\ddot{\R}$.

Using these results we get the energy conditions from (\ref{95}), (\ref{96}), (\ref{101}) and (\ref{102})  as follows: 
\ben
SEC &:&
\bar{\rho}+3\bar{p}+2\left[\alpha\R^{2}+L\dot{\phi}^2(2+\dot{\phi}^2)\right]+\frac{12\alpha\ddot{\R}}{1-\dot{\phi}^2}\left(1-\frac{2}{3}\dot{\phi}^2\right)
+\frac{12\alpha H\dot{\R}}{1-\dot{\phi}^2}\left(1-\dot{\phi}^2+\frac{2}{3}\dot{\phi}^4\right)\geq 0
\label{110}\\
NEC &:&
\bar{\rho}+\bar{p}+4\alpha\left(\ddot{\R}-H\dot{\R}\dot{\phi^2}\right)\geq 0
\label{111}\\
WEC &:&\bar{\rho}-\alpha\R^{2}-\frac{4\alpha}{1-\dot{\phi}^2}\left[\ddot{\R}\dot{\phi}^2-3\dot{\R}(1-\frac{1}{3}\dot{\phi}^4)\right]-L\dot{\phi}^2(2-\dot{\phi}^2)\geq 0
\label{112}\\
DEC &:&\bar{\rho}-\bar{p}-2\alpha\R^{2}-\frac{4\alpha}{1-\dot{\phi}^2}\Big[\ddot{\R}(1-\frac{1}{2}\dot{\phi}^2)-3H\dot{\R}(\frac{3}{2}-\frac{1}{3}\dot{\phi}^2-\frac{1}{6}\dot{\phi}^4)\Big]-2L\dot{\phi}^2(2+\dot{\phi}^2)\geq 0~~
\label{113}
\een

Again, if we put the values of $\R$, $\dot{\R}$ and $\ddot{\R}$ from (\ref{103}), (\ref{104}) and (\ref{105}) in the above equations (\ref{110}), (\ref{111}), (\ref{112}) and (\ref{113}), we easily reconstruct the energy conditions in terms of the deceleration ($q$), jerk ($j$), and snap ($s$) parameters.

\section{Discussion and conclusion}
In this work, we present a new type of modified theory, viz. $f(\R, L(X))$-gravity, with a general formalism in the context of the dark energy (using {\bf K}-essence emergent geometry) where $\R$ is the Ricci scalar of this geometry, $L(X)$ is the DBI type non-canonical Lagrangian with $X={1\over 2}g^{\mu\nu}\nabla_{\mu}\phi\nabla_{\nu}\phi$, $\phi$ is the {\bf K}-essence scalar field. The {\bf K}-essence emergent metric $\G_{\mu\nu}$ is not conformally equivalent to the gravitational metric $g_{\mu\nu}$. This new type of modified theory is a general mixing between $f(R)$ gravity and {\bf K}-essence emergent gravity based on the DBI model. 

Let us discuss some salient features of the present study which are as follows:\\
(1) It is to be noted that the modified field equation (\ref{40}) is different from the usual $f(R)$ and $f(R,L_{m})$ gravities. If we consider $f(\R,L(X))\equiv R$ and $\G_{\mu\nu}\equiv g_{\mu\nu}$ then we can easily get back to the standard Einstein field equation. The effective functional relation for the requirement of the conservation of the energy-momentum tensor is also different from the $f(R,L_{m})$-gravity. We derive the modified Friedmann equations for the $f(\R,L(X))$-gravity considering the background gravitational metric as flat FLRW and the {\bf K}-essence scalar field $\phi$ being simply a function of time only, which are quite different from the Friedmann equations of the standard $f(R)$ gravity.

(2) For the particular choice ({\it viz.} Starobinksy-type), Eq. (\ref{67}) of $f(\R,L)$ and from the requirement of the energy-momentum conservation (\ref{45}), the kinetic energy of the {\bf K}-essence scalar field is a constant. This value of $\dot\phi^{2}$($=0.888$) is less than unity which is comparable with the range of $\dot\phi^{2}$. It is also to be noted that the {\bf K}-essence theory can be used to investigate the effects of the presence of dark energy on cosmological scenarios. In this context, if we consider $\dot\phi^{2}$ be dark energy density in unit of critical density as~\cite{gm1,gm2,gm3}, then the value of dark energy density, i.e., $\dot\phi^{2}=\frac{8}{9}=0.888$ indicates that the present universe is dark energy dominated. 

It is well known that the present observational value~\cite{planck1,planck2} of dark energy density is approximately $0.75$. Therefore, we note that in the context of dark energy regime our result is in good agreement with the observational data. Now-a-days people believe that the dark energy is one of the reasons for the accelerating Universe. So, our value of dark energy density may indicate that the Universe is more accelerating. From Figs. 3 and 4, we also observe that our model is observationally verified for certain values of parameters. According to Fig. 3, the negativity of pressure can be achieved after a certain value of time which may be responsible for the accelerating universe. The variation of EOS ($\o$) with time ($t$) in Fig. 4 shows that $\o$ approaches the negative values which corresponds to observational result~\cite{Tripathi} of the present universe.

Also, we would like to put here the following two special aspects which emerge from of the present investigation:\\ 
(i) This model can open up an alternative window to explore the current cosmic acceleration without a stringent condition of invoking an exotic component as the dark energy. In other words, this theory seems interesting from a purely gravitational theory standpoint, rather than the cosmological context of dark energy whose very existence is still an issue of doubt~\cite{subir} in the context of the latest analysis of data from the Planck consortium~\cite{planck1,planck2}. However, the arbitrariness in the choice of different functional forms of $f(\R,L(X))$ based on DBI Lagrangian gives rise to the problem of how to constrain the many possible $f(\R,L(X))$ gravity theories on physical grounds. In this context, we have shed some light on this issue by discussing some constraints on general $f(\R,L(X))$ gravity from the so-called energy conditions. 

(ii) Also, we have derived the null, strong, weak and dominant energy conditions in the framework of $f(\R,L(X))$ gravity from the Raychaudhuri equations. These energy conditions are different from the usual $f(R,L_{m})$ and $f(R)$ theories. With the help of the specific form of $f(\R,L(X))$, we also have derived these energy conditions in the $f(\R,L(X))$-gravity. 

However, we intend to report on this interesting theory in the near future encompassing the multifarious aspects of Cosmology.\\

\appendix
\section{Appendix:}
\subsection{{\bf : The modified field equation in $f(R,L_{m})$-Gravity}}

Let us consider here the well known $f(R)$ and $f(R,L_{m})$ gravity, where $R$ is the Ricci scalar with respect to the gravitational metric $g_{\mu\nu}$ and $L_{m}$ is the matter Lagrangian. The total action for the $f(R)$ gravity is~\cite{sotiriou,felice}
\ben
S=\frac{1}{2\kappa}\int d^{4}x\sqrt{-g}f(R)+S_{M}(g_{\mu\nu},\psi),
\label{1}
\een
where $S_{M}$ is matter term, $\psi$ denotes the matter fields, $\k=8\pi G$, $G$ is the gravitational constant, $g$ is the
determinant of the gravitational metric and $R~(=g^{\mu\nu}R_{\mu\nu})$ is the Ricci scalar. 

Varying which with respect to the gravitational metric we achieve the modified field equation as
\ben
&&f'(R)R_{\mu\nu}-\frac{1}{2}f(R)g_{\mu\nu}-[\nabla_{\mu}\nabla_{\nu}-g_{\mu\nu}\Box]f'(R)=\k T_{\mu\nu},\nonumber\\
\label{2}
\een
with
\ben
T_{\mu\nu}= \frac{-2}{\sqrt{-g}}\frac{\delta S_{M}}{\delta g^{\mu\nu}},
\label{3}
\een
where $f'(R)=\frac{\p f(R)}{\p R}$, $\nabla_{\mu}$ is covariant derivative with respect to the gravitational metric and $\square\equiv \nabla^{\mu}\nabla_{\mu}$.

On the other hand, the action for the $f(R,L_{m})$ gravity is~\cite{harko,wang2}
\ben
S=\frac{1}{2\kappa}\int d^{4}x\sqrt{-g}f(R,L_{m}),
\label{4}
\een
where $f(R,L_{m})$ is an arbitrary function of the Ricci scalar $R$, and the Lagrangian density corresponding to matter $L_{m}$. The energy-momentum tensor is
\ben
T_{\mu\nu}= \frac{-2}{\sqrt{-g}}\frac{\delta (\sqrt{-g}L_{m})}{\delta g^{\mu\nu}}=-2\frac{\p L_{m}}{\p g^{\mu\nu}}+g_{\mu\nu}L_{m},~~\label{5}
\een
where the Lagrangian density $L_{m}$ is only matter dependent on the metric tensor components $g_{\mu\nu}$. 

The modified field equations of the $f(R,L_{m})$-gravity model is
\ben
&&f_{R}(R,L_{m})R_{\mu\nu}+(g_{\mu\nu}\square-\nabla_{\mu}\nabla_{\nu})f_{R}(R,L_{m})-\frac{1}{2}[f(R,L_{m})-L_{m}f_{L_{m}}(R,L_{m})]g_{\mu\nu}\nonumber\\&&=\frac{1}{2}f_{L_{m}}(R,L_{m})T_{\mu\nu},~~~
\label{6}
\een
where $f_{R}(R,L_{m})=\p f(R,L_{m})/\p R$ and $f_{L_{m}}(R,L_{m})=\p f(R,L_{m})/\p L_{m}$. However, if $f(R,L_{m})=R/2+L_{m}$, then the above Eq. (\ref{6}) reduces to the usual field equation $R_{\mu\nu}-(1/2)g_{\mu\nu}R=\k T_{\mu\nu}$.

\subsection{{\bf : Brief review of energy conditions in General Relativity}}
Following most of the techniques of~\cite{atazadeh,santos,capoz,wang1,wang2,berg,carroll}, we will derive the energy conditions for modified ($f(R)$, $f(R,L_{m})$, etc.) gravities. From these theories we can approach to the Null Energy Condition (NEC) and Strong Energy Condition (SEC) in the context of GR. The origin of these energy conditions comes from the Raychaudhuri equations. Let $u^{\mu}$ be the tangent vector field to a congruence of time-like geodesics in a space-time manifold endowed with a metric $g_{\mu\nu}$. Therefore, the Raychaudhuri equation~\cite{Raychaudhuri1,Raychaudhuri2,Raychaudhuri3,blau,Bhattacharyya} is
\ben
\frac{d\t}{d\tau}=-\frac{1}{3}\t^{2}-\sg_{\mu\nu}\sg^{\mu\nu}+\o_{\mu\nu}\o^{\mu\nu}-R_{\mu\nu}u^{\mu}u^{\nu},
\label{7}
\een
where $R_{\mu\nu}$ is the Ricci tensor corresponding to the metric $g_{\mu\nu}$, and $\t$, $\sg_{\mu\nu}$, and $\o_{\mu\nu}$ are the expansion, shear, and rotation associated with the congruence, respectively. While in the case of a congruence of
null geodesics defined by the vector field $k^{\mu}$, the Raychaudhuri equation~\cite{blau} is given by
\ben
\frac{d\t}{d\tau}=-\frac{1}{2}\t^{2}-\sg_{\mu\nu}\sg^{\mu\nu}+\o_{\mu\nu}\o^{\mu\nu}-R_{\mu\nu}k^{\mu}k^{\nu}.
\label{8}
\een

These equations are purely based on geometric statements, and as such it makes no reference to any gravitational field equations. In other words, the Raychaudhuri equation can be thought of as geometrical identities which do not depend on any gravitational theory. These equations are provide the evolution of the expansion of a geodesic congruence. However, since the GR field equations relate $R_{\mu\nu}$ to the energy-momentum tensor $T_{\mu\nu}$, the combination of Einstein and Raychaudhuri equations can be used to restrict energy-momentum tensors on physical ground. Indeed, the shear is a ‘‘spatial’’ tensor, given by $\sg^{2}\equiv \sg_{\mu\nu}\sg^{\mu\nu}\geq 0$. 

Thus it is clear from Raychaudhuri equation that for any hypersurface orthogonal congruences ($\o_{\mu\nu}\equiv 0$) the condition for attractive gravity (convergence of timelike geodesics or geodesic focusing) reduces to ($R_{\mu\nu}u^{\mu}u^{\nu}\geq 0$), which by virtue of Einstein’s equation implies
\ben
R_{\mu\nu}u^{\mu}u^{\nu}=(T_{\mu\nu}-\frac{T}{2}g_{\mu\nu})u^{\mu}u^{\nu}\geq 0,
\label{9}
\een
where $T$ is trace of the energy momentum tensor $T_{\mu\nu}$ ($\k=1$). Here Eq. (\ref{9}) is nothing but the SEC stated in a coordinate-invariant way in terms of $T_{\mu\nu}$ and vector fields of fixed (time-like) character. Thus, in the context of GR, the SEC ensures the fact that the gravity is attractive. In particular, for a perfect fluid of density $\rho$ and pressure $p$
\ben 
T_{\mu\nu}=(\rho +p)u_{\mu}u_{\nu}-pg_{\mu\nu}
\label{10}
\een
and the restriction given by Eq. (\ref{9}) takes the familiar form for the SEC, i.e., $\rho +3p \geq 0$.

On the other hand, the condition for the convergence (geodesic focusing) of hypersurface
orthogonal ($\o_{\mu\nu}\equiv 0$) congruences of null geodesics along
with Einstein’s equation implies
\ben
R_{\mu\nu}k^{\mu}k^{\nu}=T_{\mu\nu}k^{\mu}k^{\nu} \geq 0
\label{11}
\een
which is the condition for NEC written in a coordinate-invariant way.

Thus, in GR the NEC ultimately encodes the null geodesic focusing due to the gravitational attraction. For the energy-momentum tensor of a perfect fluid (\ref{10}) the above condition (\ref{11}) reduces to the well-known form of the NEC, i.e., $\rho +p\geq 0$.

The Weak Energy Condition (WEC) states that $T_{\mu\nu}u^{\mu}u^{\nu}\geq 0$ for all time-like vectors $u^{\mu}$, or equivalently for perfect fluid it is $\rho>0$ and $\rho +p>0$. The Dominant Energy Condition (DEC) includes the WEC as well as the additional requirement that $T_{\mu\nu}u^{\mu}$ is a non space-like vector i.e., $T_{\mu\nu}T^{\nu}_{\l}u^{\mu}u^{\l}\leq 0$. For a perfect fluid, these conditions together are equivalent to the simple requirement that $\rho\geq |p|$, the energy density must be nonnegative, and greater than or equal to the magnitude of the pressure.\\

{\it In $f(R)$-gravity}~\cite{atazadeh,santos}, the energy conditions for perfect fluid are given by
\ben
&&SEC:~~\rho +3p-f+Rf'+3(\ddot{R}+\dot{R}H)f''+3\dot{R}^{2}f'''\geq 0,\nonumber\\
&&NEC:~~\rho +p+(\ddot{R}-\dot{R}H)f''+\dot{R}^{2}f'''\geq 0,\nonumber\\
&&WEC:~~\rho+\frac{1}{2}(f-Rf')-3\dot{R}Hf''\geq 0,\nonumber\\
&&DEC:~\rho-p+f-Rf'-(\ddot{R}+5\dot{R}H)f''-\dot{R}^{2}f'''\geq 0,\nonumber\\
\label{12}
\een
where $f'=\frac{\p f(R)}{\p R}$.\\

{\it In $f(R,L_{m})$-gravity}~\cite{wang2}, the energy conditions are
\ben
SEC&:&~~\rho +3p-\frac{2}{f_{L_{m}}}[f-Rf']+\frac{6}{f_{L_{m}}}[\dot{R}^{2}f''' +\ddot{R}f'' +H\dot{R}f'']-2L_{m}\geq 0.\nonumber\\
NEC&:&~~\rho +p+\frac{2}{f_{L_{m}}}[\dot{R}^{2}f''' +\ddot{R}f'' ]\geq 0,\nonumber\\
WEC&:&~~\rho+\frac{1}{f_{L_{m}}}[f-Rf']-\frac{6}{f_{L_{m}}}H\dot{R}f''+L_{m}\geq 0,\nonumber\\
DEC&:&~~\rho-p+\frac{2}{f_{L_{m}}}[f-Rf']-\frac{2}{f_{L_{m}}}\Big[\dot{R}^{2}f''' +\ddot{R}f''+6H\dot{R}f''\Big]+2L_{m}\geq 0,
\label{13}
\een
where $f'=\frac{\p f(R,L_{m})}{\p R}$.

\section*{Acknowledgement}
G.M. and A.P. acknowledge the DSTB, Government of West Bengal, India for financial support through the Grants No.: 322(Sanc.)/ST/P/S\&T/16G-3/2018 dated 06.03.2019. Also, the authors would like to thank the referee for illuminating suggestions to improve the manuscript.


\vspace{1in}
\begin{thebibliography}{99}

\bibitem{weyl} H. Weyl, {\it A new extension of relativity theory},  Ann. Phys. {\bf 59}, 101 (1919).

\bibitem{eddin} A.S. Eddington, {\it The mathematical theory of relativity}, Cambridge University Press, Cambridge (1923).

\bibitem{uti} R. Utiyama, B.S. DeWitt, {\it Renormalization of a classical gravitational field interacting with quantized matter fields}, J. Math. Phys. {\bf 3}, 608 (1962).

\bibitem{birrel} N.D. Birrell, P.C.W. Davies, {\it Quantum fields in curved spacetime}, Cambridge University Press, Cambridge (1982).

\bibitem{staro} A.A. Starobinsky, {\it A new type of isotropic cosmological models without singularity}, Phys. Lett. B {\bf 91}, 9 (1980).

\bibitem{sotiriou} T.P. Sotiriou, V. Faraoni, {\it $f(R)$ theories of gravity}, Rev. of  Mod. Phys. {\bf 82}, 451 (2010).

\bibitem{felice} A. De Felice, S. Tsujikawa, {\it $f(R)$ theories}, Living Rev. Relativity {\bf 13}, 3 (2010).

\bibitem{dunsby} P.K.S. Dunsby et al., {\it $\lambda$CDM universe in $f(R)$ gravity}, Phys. Rev. D {\bf 82}, 023519 (2010).

\bibitem{mukherjee} A. Mukherjee, N. Banerjee, {\it Acceleration of the universe in f(R) gravity models}, Astrophys. Space Sci. {\bf 352}, 893 (2014).

\bibitem{atazadeh} K. Atazadeh et al., {\it Energy conditions in $f(R)$ gravity and Brans-Dicke theories}, Int. J. Mod. Phys. D {\bf 18}, 1101 (2009). 

\bibitem{santos} J. Santos et al., {\it Energy conditions in $f(R)$ gravity}, Phys. Rev. D {\bf 76}, 083513 (2007).

\bibitem{capoz} S. Capozziello et al., {\it The role of energy conditions in $f(R)$ cosmology}, Phys. Lett. B {\bf 781}, 99 (2018). 

\bibitem{wang1} J. Wang et al., {\it Energy conditions and stability in generalized $f(R)$ gravity with arbitrary coupling between matter and geometry}, Phys. Lett. B {\bf 689}, 133 (2010).

\bibitem{berg} S.E. Perez Bergliaffa, {\it Phantom and Quintessence fields coupled to scalar curvature in general $f(R)$ gravity theory}, Phys. Lett. B {\bf 642},  311 (2006).  

\bibitem{alba1} F.D. Albareti et al., {\it The Raychaudhuri equation in homogeneous cosmologies}, JCAP {\bf 03}, 012 (2014).

\bibitem{alba2} F.D. Albareti et al., {\it On the non-attractive character of gravity in f(R) theories}, JCAP {\bf 07}, 009 (2013).

\bibitem{krori} K.D. Krori et al., {\it Raychaudhuri equation, big bang and accelerating universe}, Ind. J. Phys. {\bf 82(5)}, 531 (2008).

\bibitem{harko} T. Harko, F.S.N. Lobo, {\it $f(R,L_m)$ gravity}, Eur. Phys. J. C  {\bf 70}, 373, (2010).

\bibitem{wang2} J. Wang, K. Liao, {\it Energy conditions in $f(R, L_m)$ gravity}, Class. Quantum Gravit., {\bf 29}, 215016 (2012). 

\bibitem{Goheer1} N. Goheer, R. Goswami P. Dunsby, K. Ananda, {\it Coexistence of matter dominated and accelerating solutions in 
$f(G)$ gravity}, Phys. Rev. D {\bf 79}, 121301(R) (2009).

\bibitem{Goheer2} N. Goheer, J. Larena, P.K.S. Dunsby, {\it Power-law cosmic expansion in $f(R)$ gravity models}, Phys. Rev. D {\bf 80}, 061301(R) (2009).

\bibitem{Singh} C.P. Singh, V. Singh, {\it Power-law expansion and scalar field cosmology in higher derivative theory}, Int. J. Theor. Phys. {\bf 51}, 1889 (2012).


\bibitem{Harko3}
T. Harko et. al., ``f(R, T ) gravity”, Phys. Rev. D {\bf 84}, 024020, (2011).


\bibitem{Picon1} C. Armendariz-Picon et al., {\it Essentials of {\bf K-}essence}, Phys. Rev. {\bf D 63}, 103510, (2001)

\bibitem{Picon2} C. Armendariz-Picon et al., {\it Dynamical solution to the problem of a small cosmological constant and late-time cosmic acceleration}, Phys. Rev. Lett. {\bf 85}, 4438 (2000)

\bibitem{babi1} M. Visser, C. Barcelo, S. Liberati, {\it Analogue models of and for gravity}, Gen. Relativ. Gravit. {\bf 34} 1719 (2002).

\bibitem{babi2} E. Babichev, V. Mukhanov, A. Vikman, {\it Escaping from the black hole?}, JHEP {\bf 0609}, 061 (2006).

\bibitem{babi3} E. Babichev, V. Mukhanov, A. Vikman, {\it {\bf K-} essence, superluminal propagation, causality and emergent geometry}, JHEP {\bf 0802} 101 (2008).

\bibitem{babi4} A. Vikman, {\it K-essence: Cosmology, causality and Emergent Geometry},  Dissertation an der Fakultat fur Physik, Arnold Sommerfeld Center for Theoretical Physics, der Ludwig-Maximilians-Universitat Munchen, Munchen, den 29.08.2007.

\bibitem{babi5} E. Babichev, V. Mukhanov, A. Vikman, {\it Looking beyond the Horizon}, WSPC-Proceedings (October 23, 2008).

\bibitem{scherrer1} R.J. Scherrer, {\it Purely kinetic {\bf K-}essence as unified dark matter}, Phys. Rev. Lett. {\bf 93} 011301 (2004).

\bibitem{scherrer2} L.P. Chimento, {\it Extended tachyon field, Chaplygin gas, and solvable K-essence cosmologies}, Phys. Rev. D {\bf 69} 123517 (2004).

\bibitem{born1} M. Born, L. Infeld, {\it Foundations of the new field theory}, Proc. Roy. Soc. Lond A {\bf 144}, 425 (1934).

\bibitem{born2} W. Heisenberg, {\it On the theory of explosive showers in cosmic rays}, Zeit. Phys. {\bf 113} 61 (1939).

\bibitem{born3} P.A.M. Dirac, {\it An extensible model of the electron}, Proc. R. Soc. Lond. A {\bf 268}, 57 (1962).



\bibitem{gm1} D. Gangopadhyay, G. Manna, {\it The Hawking temperature in the context of dark energy}, Eur. Phys. Lett. {\bf 100} 49001 (2012).

\bibitem{gm2} G. Manna, D. Gangopadhyay, {\it The Hawking temperature in the context of dark energy for Reissner–Nordstr{\"o}m and Kerr background}, Eur. Phys. J. C {\bf 74} 2811 (2014).

\bibitem{gm3} G. Manna, B. Majumder, {\it The Hawking temperature in the context of dark energy for Kerr–Newman and Kerr–Newman–AdS backgrounds}, Eur. Phys. J. C {\bf 79}, 553 (2019).


\bibitem{Mukohyama} Shinji Mukohyama et. al. {\it Is the DBI scalar field as fragile as other k-essence fields?}, Phys. Rev. D.{\bf 94}, 023514 (2016)


\bibitem{nojiri} S. Nojiri et. al., {\it {\bf K-}essence $f(R)$ gravity inflation},  Nuc. Phys. B {\bf 941}, 11 (2019).

\bibitem{odin} S.D. Odintsov et. al., {\it $f(R)$ gravity {\bf K-}essence late-time phenomenology}, Phys. Dark Univ., {\bf 29},100563, (2020)

\bibitem{oiko} V.K. Oikonomou, N. Chatzarakis, {\it The phase space of {\bf K-}essence $f(R)$ gravity theory}, Nuc. Phys. B {\bf 956}, 115023 (2020).

\bibitem{Bahcall} N. Bahcall et al., {\it The cosmic triangle: Revealing the state of the universe}, Science {\bf 284}, 1481  (1999) [and references therein].

\bibitem{Kang} Jin U Kang et al., {\it Attractor scenarios and superluminal signals in {\bf K-}essence cosmology}, Phys. Rev. {\bf D.76}, 083511 (2007)

\bibitem{Caldwell} R.R. Caldwell et al., {\it Cosmological imprint of an energy component with general equation of state}, Phys. Rev. Lett. {\bf 80}, 1582 (1998).

\bibitem{Frieman} J. Frieman et al., {\it Cosmology with ultralight pseudo Nambu-Goldstone bosons}, Phys. Rev. Lett. {\bf 75}, 2077, (1995).

\bibitem{Peebles} P.J.E. Peebles, B. Ratra, {\it Cosmology with a time-variable cosmological constant}, Astrophys. J. Lett. {\bf 325}, L17 (1988).

\bibitem{Ratra} B. Ratra, P.J.E. Peebles, {\it Cosmological consequences of a rolling homogeneous scalar field}, Phys. Rev. D {\bf 37}, 3406 (1988).

\bibitem {Zlatev1} I. Zlatev, P.J. Steinhardt, {\it A tracker solution to the cold dark matter cosmic coincidence problem}, Phys. Lett. B {\bf 459}, 570 (1999).

\bibitem{Zlatev2} P.J. Steinhardt, L. Wang, I. Zlatev, {\it Cosmological tracking solutions}, Phys. Rev. {\bf D59}, 123504 (1999).

\bibitem{Zlatev3} I. Zlatev, Limin Wang, P. J. Steinhardt, {\it Quintessence, cosmic coincidence, and the cosmological constant}, Phys. Rev. Lett. {\bf 82}, 896 (1999)

\bibitem{eric} J.K. Erickson et al., {\it Measuring the speed of sound of quintessence'}, Phys. Rev. Lett. {\bf 88}, 121301, (2002).

\bibitem{dedeo} S. DeDeo, R.R. Caldwell, P.J. Steinhardt, {\it Effects of the sound speed of quintessence on the microwave background and large scale structure}, Phys. Rev. D {\bf 67}, 103509 (2003).

\bibitem{bean} R. Bean, O. Dore, {\it Probing dark energy perturbations: the dark energy equation of state and speed of sound as measured by WMAP}, Phys. Rev. D {\bf 69}, 083503 (2004).

\bibitem{Bonvin} C. Bonvin et. al., {\it No-Go Theorem for k-Essence Dark Energy}, Phys. Rev. Lett. {\bf 97}, 081303, (2006).

\bibitem{Yang} R. Yang et al., {\it The evolution of the power law {\bf K-}essence cosmology}, Astrophys. Space Sci. {\bf 356}, 399 (2014).

\bibitem{Sawicki} I. Sawicki et. al., {\it Seeding supermassive black holes with a nonvortical dark-matter subcomponent}, 
Phys. Rev. D {\bf 88}, 083520, (2013).

\bibitem{Kunz} M. Kunz et. al., {\it Using dark energy to suppress power at small scales}, Phys. Rev. D {\bf 92}, 063006, (2015).

\bibitem{Linde} A.D. Linde, {\it A new inflationary universe scenario: a possible solution of the horizon, flatness, homogeneity, isotropy and primordial monopole problems}, Phys. Lett. B {\bf 108}, 389 (1982). 

\bibitem{Albrecht} A. Albrecht, P.J. Steinhardt, {\it Cosmology for grand unified theories with radiatively induced symmetry breaking}, Phys. Rev. Lett. {\bf 48}, 1220 (1982).

\bibitem{Dvali} G. Dvali, S.-H. Henry Tye, {\it Brane inflation}, Phys. Lett. B {\bf 450}, 72 (1999).

\bibitem{Kachru} S. Kachru et al., {\it Towards inflation in string theory}, JCAP {\bf 0310}, 013 (2003).

\bibitem{Alishahiha} M. Alishahiha, E. Silverstein, D. Tong, {\it DBI in the sky: non-Gaussianity from inflation with a speed limit}, Phys. Rev. D {\bf 70}, 123505 (2004).

\bibitem{Silverstein1} E. Silverstein, D. Tong, {\it Scalar speed limits and cosmology: Acceleration from Deceleration}, Phys. Rev. D {\bf 70}, 103505 (2004).

\bibitem{Chen1} X. Chen, {\it Multithroat brane inflation}, Phys. Rev. D {\bf 71}, 063506 (2005).

\bibitem{Weinberg} S. Weinberg, {\it Effective field theory for inflation},  Phys. Rev. D {\bf 77}, 123541 (2008).

\bibitem{Chen2} X. Chen et. al., {\it Observational signatures and non-Gaussianities of general single-field inflation}, JCAP {\bf 0701}, 002 (2007)

\bibitem{Panda}
Panda et, al., ``$f(\bar{R},T)-$gravity in the context of dark energy", arXiv. https://doi.org/10.48550/arXiv.2206.14808 (2022) 

\bibitem{wald} R.M. Wald, {\it General Relativity}, Univ. Chicago Press (1984).

\bibitem{misner} C. Misner, K.S. Thorne, J. Wheeler, {\it Gravitation}, W.H. Freeman and Company (1970).

\bibitem{koiv} T. Koivisto, {\it A note on covariant conservation of energy–momentum in modified gravities},  Class. Quantum Gravit. {\bf 23} 4289 (2006).

\bibitem{harko1} T. Harko, {\it Modified gravity with arbitrary coupling between matter and geometry}, Phys. Lett. B {\bf 669}, 376 (2008). 

\bibitem{gm4} D. Gangopadhyay, G. Manna, {\it Cosmology in presence of dark energy in an emergent gravity scenario}, arXiv:gr-qc/1502.06255 (2015).

\bibitem{Raychaudhuri1} A. Raychaudhuri, {\it Relativistic cosmology. I}, Phys. Rev. {\bf 98}, 1123 (1955).

\bibitem{Raychaudhuri2} A. Raychaudhuri, {\it Relativistic and Newtonian cosmology},  Z. Astrophysik, {\bf 43}, 161 (1957).

\bibitem{Raychaudhuri3} A. Raychaudhuri, {\it Singular state in relativistic cosmology}, Phys. Rev. {\bf 106}, 172 (1957).

\bibitem{blau} M. Blau, {\it Lecture Notes on General Relativity}, http://www.blau.itp.unibe.ch/GRLecturenotes.html (2020).

\bibitem{Bhattacharyya} I. Bhattacharyya, S. Ray, {\it A generalized form of the Raychaudhuri equation}, Int. J. Mod. Phys. D, {\bf 30}, 2150092 (2021) 

\bibitem{carroll} S. Carroll, {\it Spacetime and Geometry: An Introduction to General Relativity}, Addison Wesley, New York (2004). 

\bibitem{harr} E.R. Harrison, {\it Observational tests in cosmology}, Nature (London) {\bf 260}, 591 (1976).

\bibitem{land} P. Landsberg, {\it Q in cosmology}, Nature (London) {\bf 263}, 217 (1976).

\bibitem{visser1} M. Visser, {\it Jerk, snap and the cosmological equation of state}, Class. Quantum Gravit. {\bf 21}, 2603 (2004).

\bibitem{visser2} M. Visser, {\it Cosmography: Cosmology without the Einstein equations}, Gen. Relativ. Gravit. {\bf 37}, 1541 (2005).

\bibitem{subir} J.T. Nielsen, A Guffanti, S. Sarkar, {\it Marginal evidence for cosmic acceleration from Type Ia supernovae}, Sci. Rep. {\bf 6} 3559 (2016).

\bibitem{planck1} P.R. Ade et al., {\it Planck 2015 results. XIII. Cosmological parameters}, Astron. Astrophys. {\bf 594}, A13 (2016).

\bibitem{planck2} N. Aghanim et al.,  {\it Planck 2018 results. VI. Cosmological parameters, Planck Collaboration}, Astron. Astrophys. {\bf 641}, A6 (2020).

\bibitem{Kehagias} A. Kehagias et al., {\it Remarks on the Starobinsky model of inflation and its descendants}, Phys. Rev. D {\bf 89} 043527 (2013).

\bibitem{Tripathi} A. Tripathi et. al., {\it Dark energy equation of state parameter and its evolution at low redshift}, JCAP {\bf 06}, 12 (2017)

\bibitem{Moraes} P.H.R.S. Moraes et. al. {\it A cosmological scenario from the Starobinsky model within the $f(R,T)$ Formalism}, Adv. Astron. {\bf 8574798} (2019).

\end{thebibliography}
\end{document}